# Unlocking Criminal Hierarchies: A Survey, Experimental, and Comparative Exploration of Techniques for Identifying Leaders within Criminal Networks

Kamal Taha[a*], Abdulhadi Shoufan[b], and Aya Taha[c]

[a] *Department of Computer Science, Khalifa University, Abu Dhabi, UAE, kamal.taha@ku.ac.ae*
[b] *Department of Computer Science, Khalifa University, Abu Dhabi, UAE, abdulhadi.shoufan@ku.ac.ae*
[c] *Brighton College, Dubai, United Arab Emirates, Email: bc004814@brightoncollegedubai.ae*

**Abstract**

This survey paper offers a thorough analysis of techniques and algorithms used in the identification of crime leaders within criminal networks. For each technique, the paper examines its effectiveness, limitations, potential for improvement, and future prospects. The main challenge faced by existing survey papers focusing on algorithms for identifying crime leaders and predicting crimes is effectively categorizing these algorithms. To address this limitation, this paper proposes a new methodological taxonomy that hierarchically classifies algorithms into more detailed categories and specific techniques. The paper includes empirical and experimental evaluations to rank the different techniques. The combination of the methodological taxonomy, empirical evaluations, and experimental comparisons allows for a nuanced and comprehensive understanding of the techniques and algorithms for identifying crime leaders, assisting researchers in making informed decisions. Moreover, the paper offers valuable insights into the future prospects of techniques for identifying crime leaders, emphasizing potential advancements and opportunities for further research. Here's an overview of our empirical analysis findings and experimental insights, along with the solution we've devised: (1) PageRank and Eigenvector centrality are reliable for mapping network connections, (2) Katz Centrality can effectively identify influential criminals through indirect links, stressing their significance in criminal networks, (3) current models fail to account for the specific impacts of criminal influence levels, the importance of socio-economic context, and the dynamic nature of criminal networks and hierarchies, and (4) we propose enhancements, such as incorporating temporal dynamics and sentiment analysis to reflect the fluidity of criminal activities and relationships, which could improve the detection of key criminal figures as their roles or tactics evolve.

## 1. Introduction

Due to societal progress, organized crime has emerged as the primary type of criminal structure. Criminal organizations now operate within intricate social networks, making it challenging to distinguish between innocent individuals and members involved in criminal activities due to limited data availability [1]. In real-world investigations, while some conspirators are known and others are not, the goal is to ascertain the involvement of uncertain members and pinpoint the leaders prior to making arrests.

In criminal investigations, the challenge lies in mapping the criminal network's structure to identify its leaders and participants before proceeding with arrests. Criminal groups often share similarities through friend-of-a-friend relationships, co-offending experiences, referral chains, and the need for specialized expertise [2]. Connections between individuals can be categorized as strong or weak ties, representing different levels of interaction. Strong ties are close and trusted relationships, while weak ties are more distant, like co-workers [3].

Criminal groups often share similarities through friend-of-a-friend positions, co-offending experiences, referral chains, and the need for specialized expertise [4]. Individual connections range from strong, trust-based relationships with family and friends, to weaker ties like those with acquaintances or colleagues. Both strong and weak ties have their own pros and cons [3]. Crime data is classified by crime type [5]. Analyzing crime by category aids in crime prevention and reduction. Organizational structures and common locations affect crime frequency. Studying crime patterns over time highlights hotspots and helps predict and reduce future incidents. In-depth analysis of structured crime data enhances our grasp of criminal activities, with historical records pinpointing areas of concern.

Within criminal networks, brokers play an even more critical role due to the absence of formal regulations and mechanisms governing transactions and conflicts in stateless environments [6, 7, 8, 9]. Success within criminal organizations relies heavily on social connections that provide access to profitable opportunities [6, 10]. In contemporary times, criminals must strike a delicate balance between efficiently managing illicit activities and ensuring the security of the group [11]. Criminal leaders act as brokers in their networks, with higher betweenness centrality scores indicating their strategic role [12].

In organized crimes, leaders serve as bridges between criminals, individuals in businesses, and politics, exploiting these connections for criminal opportunities [6, 13, 14, 15]. Identifying criminal leaders through wiretap data is limited due to cautiousness and minimized telecommunications usage by criminals [16, 17, 18]. Balancing efficiency and security, criminals limit information sharing to avoid detection, with leaders using telecommunications sparingly [19, 11]. Leaders may delegate risky activities to middle-level criminals.

### 1.1. Motivations and Key Contributions

1) *Main Challenge and Proposed Solution*
   a) <u>Current Issue:</u> Survey papers in the field of algorithms for identifying crime leaders and predicting crimes struggle with effectively categorizing these algorithms. They often use broad and non-specific groupings. This lack of specificity can lead to confusion when classifying unrelated algorithms and result in inaccurate evaluations using the same metrics.
   b) <u>Proposed Solution:</u> This paper introduces a new methodological taxonomy. It hierarchically classifies algorithms for crime leaders prediction into specific and detailed categories and techniques, enabling a precise and systematic approach to categorization.
2) *Comprehensive Survey and Enhanced Assessment*
   a) <u>Survey Goals:</u> We provide a survey of algorithms, focusing on those that use *same* sub-techniques, techniques, sub-categories, and categories.
   b) <u>Benefits of the Taxonomy:</u> Utilizing this taxonomy allows for more accurate assessments and comparisons of algorithms. This leads to a deeper understanding of their strengths and weaknesses and paves the way for future research.
3) *Empirical and Experimental Evaluations*
   a) <u>Empirical Evaluation:</u> The paper includes an empirical evaluation, examining various techniques for identifying crime leaders based on four distinct criteria.
   b) <u>Experimental Evaluation:</u> Through experimental evaluation, this study ranks algorithms, including those that utilize the same sub-technique, different sub-techniques within the same technique, different techniques within the same sub-category, different sub-categories within the same category, and categories.
4) *Overall Contributions*
   a) <u>Comprehensive Understanding:</u> The integration of the methodological taxonomy with empirical and experimental evaluations offers researchers a thorough and nuanced understanding of available algorithms.
   b) <u>Informed Decision-Making:</u> This approach aids researchers in making well-informed decisions about selecting appropriate techniques for specific needs.

### 1.2. Proposed Methodology-Based Taxonomy

We categorize crime leaders' identification algorithms into three main

---

[*] Corresponding author:
*Email Address:* kamal.taha@ku.ac.ae (K. Taha)

classes based on the techniques they use. These three broad classes are topology-based, clustering-based, and agent-based methods. Each of these methods is further subdivided into three tiers, with each tier being more specific than the previous one. Our methodology-based taxonomy is structured hierarchically as follows:

**Methodology category → methodology sub-category → methodology techniques → methodology sub-techniques**.

This hierarchy allows us to identify specific techniques or sub-techniques in the final level. Fig. 1 shows our methodology-based taxonomy. Our taxonomy offers the following benefits:
- *Enhanced organization:* It offers a well-organized framework for presenting the survey results. By grouping related approaches, the hierarchical structure helps readers to follow the paper's logical flow.
- *Comprehensive coverage:* The taxonomy provides thorough coverage of all pertinent methods, and its hierarchical design helps identify research gaps and areas needing more exploration.
- *Comparison of techniques:* The taxonomy aids in comparing research techniques by grouping similar methods and highlighting their similarities and differences, allowing for an assessment of their strengths and weaknesses.
- *Improved reproducibility:* The taxonomy enhances research reproducibility by clearly describing approaches, making it easier for other researchers to replicate and build upon its findings.

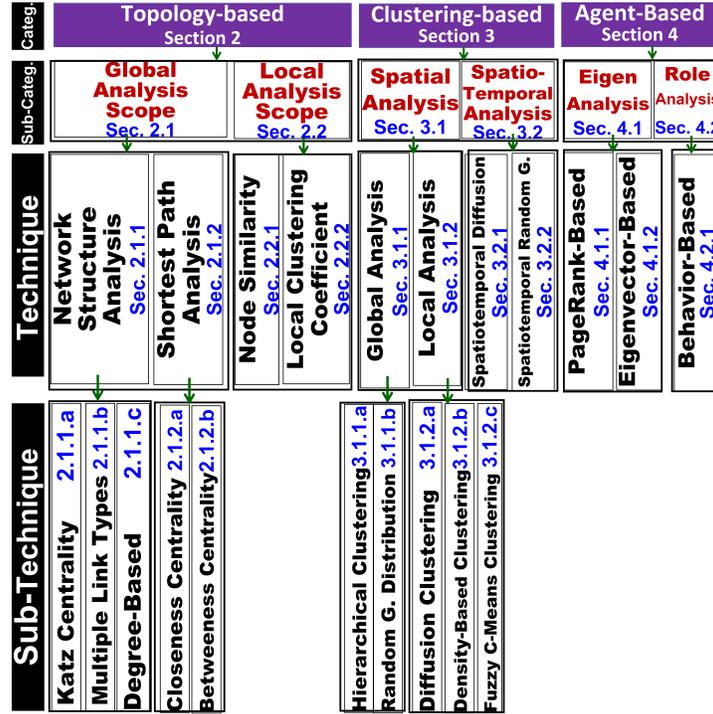

**Fig. 1:** Our methodology-based taxonomy that categorizes the algorithms for the identification of influential criminals and crime leaders into fine-grained classes in a hierarchical manner, as follows: methodology category → methodology sub-category → methodology technique → methodology sub-technique. For each category, sub-category, technique, and sub-technique, the figure also shows the section number in the manuscript that discusses it.

## 2. Topology-Based Analysis

### 2.1 Global Analysis Scope

Global topology analysis refers to the examination of the overall structure and properties of a criminal social network. Global topology analysis is a valuable approach for identifying influential individuals in a criminal social network. By examining the structural properties of the network, such as node centrality and community detection, we can gain insights into the key players and their roles within the criminal organization.

#### 2.1.1 Network-Based Model Analysis Approach

**a) Katz Centrality-Based Model**

Katz centrality emerges as an essential metric for the identification of influential figures within criminal networks, facilitating the detection of crime leaders by examining the web of direct and indirect relationships within these networks. This approach employs an adjacency matrix to capture the connections between individuals, highlighting both the presence and intensity of these links. The Katz centrality score for each node, representing individuals within the network, is determined through a formula: $C(v) = \alpha \sum(A(u, v) * C(u)) + \beta$. In this equation, $C(v)$ is the Katz centrality score of node $v$, illustrating the node's level of influence within the network. $A(u, v)$ refers to the adjacency matrix's element at row u and column v, denoting the connection's strength between nodes u and v, while $C(u)$ stands for the centrality score of node u. Parameters $\alpha$ and $\beta$ are meticulously chosen, where $\alpha$ adjusts the significance given to connections based on their path length, and $\beta$ sets a baseline score for each node.

The process involves iterating this equation until the centrality scores converge to a stable value, thereby updating the influence scores of each individual in the network. Following the computation, individuals are ranked according to their Katz centrality scores, allowing for the strategic identification of the network's most dominant figures. A specific threshold is set as a benchmark to filter out individuals exceeding a predetermined level of influence, effectively isolating key operatives within criminal circles. This methodology not only quantifies the degree of influence wielded by each member but also underscores the intricate structure of criminal networks, guiding law enforcement in targeting and dismantling the leadership of such illicit operations. Fig. 2 presents the Katz procedure.

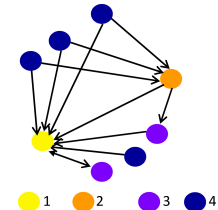

**Fig. 2:** The procedure of Katz centrality. Each node's legend number indicates the centrality rank of the node.

*i) The Rationale Behind the Usage of the Technique*

In a criminal network, individuals don't always interact directly with everyone else, but their influence can still spread through connections. Katz centrality considers both direct and indirect connections to measure influence in the network. Criminal networks often have hierarchies or chains of command, where top-ranking individuals control lower-ranking members. Katz centrality captures influence propagation by assigning higher scores to nodes connected to highly influential nodes. It considers multiple paths, network structure and varying levels of influence.

*ii) The Conditions for the Technique's Optimal Performance*

The technique can be improved by: (1) including relevant nodes beyond the main criminals to analyze the influence within the network comprehensively, such as their associates, subordinates, and other individuals involved in criminal activities, (2) accurately representing the direction of relationships in the network to capture the flow of influence correctly, (3) adapting the analysis to align with the specific characteristics of the criminal context, (4) experimenting with different damping factor values to emphasize various levels of influence and accurately capture the dynamics of the network.

*iii) Research Papers that have Employed the Technique*

Cavallaro et al. [20] used Katz centrality to identify influential individuals in Sicilian Mafia gangs. They aimed to understand the gangs' structure and organization using real-world datasets, focusing on their resilience to law enforcement. Two networks were created: one from phone call data and another from records of physical meetings within the gangs. Zhang et al. [21] integrated Katz centrality and betweenness centrality to evaluate node significance in networks, including criminal networks. This approach addressed limitations by combining the calculation of shortest paths using betweenness centrality with assigning varying weights to all paths using Katz centrality. This provided a more comprehensive measure of node importance. Calderoni et al. [22] found that Katz scores effectively utilized the entire graph structure and produced accurate results, even with limited network connectivity. They conducted experiments on networks based on meetings and recorded telephone calls among criminals.

**Table 1:** Evaluating research papers that have employed Katz Centrality

| Paper/Year | Dataset | Scalability | Interpretability | Accuracy | Efficiency |
|---|---|---|---|---|---|
| [20] 2020 | Italian mafia's calls and meetings | Acceptable | Fair | Good | Acceptable |
| [21] 2015 | Author collaboration network | Unsatisfactory | Fair | Acceptable | Fair |
| [22] 2020 | Italian crime case | Unsatisfactory | Acceptable | Good | Acceptable |

**b) Multiple Link Types Model**

To effectively identify and understand the roles of influential criminals within a criminal network, a multifaceted approach that examines diverse connections and exchanges is essential. This method starts by characterizing various types of interactions, such as drug exchanges or communication patterns, within the criminal social network. By assigning weights or strengths to these connections based on criteria like intensity, frequency, or significance, a comprehensive understanding of the network's structure and the strategic roles of its members is developed.

The analysis employs a multiple link types model that categorizes different connection types, allowing for a multilayer network analysis where each layer represents a unique type of interaction. Through this detailed examination, both direct and indirect connections are considered, enabling the identification of influential criminals whose impact is significant across various facets of the network. By applying centrality measures, importance scores are assigned to individuals based on their involvement and influence within these diverse connections.

This approach not only illuminates the complex structure of criminal networks but also highlights the critical roles of key individuals. Their influence extends beyond direct interactions to encompass a broader range of activities within the network, revealing the multi-dimensional nature of criminal hierarchies and interactions. By understanding these dynamics, law enforcement and researchers can more effectively target and dismantle criminal organizations, focusing on those individuals who play pivotal roles in maintaining and expanding these networks.

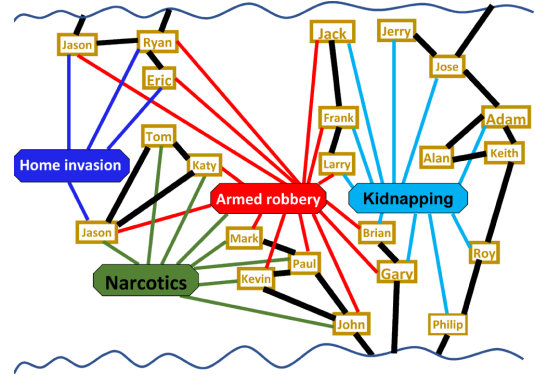

**Fig. 3:** A faction of a hypothetical criminal network with multi-link types

*i) The Rationale Behind the Usage of the Technique*

Criminal activities involve complex relationships and interactions. Considering multiple types of connections provides a more comprehensive understanding of network dynamics and individual roles. In criminal networks, interactions like partnerships, hierarchies, resource sharing, and communication patterns are present. Each connection type represents a unique aspect of these interactions. Examining different link types reveals connections that may not be apparent when focusing on a single relationship type.

*ii) The Conditions for the Technique's Optimal Performance*

The technique can be enhanced by: (1) understanding and representing each link type's specific meanings and implications in the network data (differentiate between link types by using varied edge labels, weights, or attributes), (2) evaluating the importance and significance of each link type concerning its potential influence on the network, (3) assigning suitable weights or strengths to each link type based on their relevance and impact on influence within the network, (4) utilizing algorithms capable of effectively handling multiple link types.

*iii) Research Papers that have Employed the Technique*

Bright et al. [17] studied different link types in a drug manufacturing and trafficking network. They found eight link types associated with specific resource exchanges, such as drugs. Examining multiple link types helped them understand the network structure and individuals' strategic roles. Ficara et al. [23] created a multilayer network by adding a third layer based on criminal activities committed collectively. The network had 226 actors, 454 edges within layers, and 3 layers: Meetings, Phone Calls, and Crimes. Analyzing actor and layer measures helped assess significance and dissimilarities. The multilayer approach revealed important actors not evident in separate layer examination. Schwartz and Rouselle [24] suggested choosing the indirect path with the highest indirect connection score in networks with multiple links. They showed that an influential actor's impact can be maximized through indirect connections involving multiple links. Maulana and Emmerich [25] proposed a method to examine network centrality in multiplex networks. They calculated Pareto fronts of node centrality, with each layer maximizing its own centrality. Dominance rank within a multiplex network indicated a node's significance.

**Table 2:** Evaluating papers that employed the Multiple Link Types Model

| Paper/Year | Dataset | Scalability | Interpretability | Accuracy | Efficiency |
|---|---|---|---|---|---|
| [17] 2015 | Drug trafficking network | Fair | Acceptable | Good | Fair |
| [23] 2022 | Montagna operation | Unsatisfactory | Acceptable | Acceptable | Unsatisfactory |
| [24] 2009 | Author Collected | Unsatisfactory | Unsatisfactory | Acceptable | Fair |
| [25] 2017 | Economic dataset | Unsatisfactory | Fair | Good | Fair |

## c) Degree-Based Analysis

Degree-based analysis is a pivotal approach for identifying influential figures within criminal networks. It meticulously examines the connectivity or centrality of nodes, which represent individuals, by focusing on their degree of connectivity. The degree of a node is determined by its number of connections or links to other nodes within the network. The degree centrality $C(v)$ of a node $v$ is given by: $C(v) = deg(v)/(N-1)$, where $deg(v)$ is the degree of the node $v$ (the number of edges incident to $v$), $N$ is the total number of nodes in the graph, and $N-1$ is the maximum possible degree of a node in an undirected graph. This method is instrumental in pinpointing individuals with a high degree of connectivity, who are deemed influential due to their extensive network. These individuals are key players in the dissemination of information, resources, or criminal activities, and their high degree of connectivity signifies their control over the flow of information and criminal endeavors.

Also, degree-based analysis not only highlights these central figures but also aids in understanding the network's structure. By analyzing the degree distribution, it's possible to discern if the network exhibits a specific pattern, such as a power-law distribution, indicating that a few individuals hold significantly more connections than the rest. Identifying these highly connected individuals is crucial, as they often assume critical roles in the operation of the criminal network. Targeting these high-degree nodes can lead to a significant disruption in the network's functionality and dynamics, underscoring the effectiveness of degree-based analysis in dismantling criminal networks and detecting crime leaders.

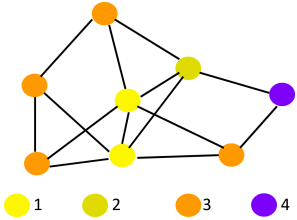

**Fig. 4:** Illustration of the Degree-based analysis procedure. Each node's legend number indicates the centrality rank of the node

### i) The Rationale Behind the Usage of the Technique

Criminal networks have complex social structures with diverse relationships and interactions. Analyzing degrees helps understand the network's structure by quantifying connections and identifying key players in central positions. Information flow is crucial in criminal networks for coordination, resource sharing, and control. Individuals with high degrees of connections have access to extensive information and engage in multiple interactions. Examining their degree centrality reveals influential individuals who facilitate information dissemination.

### ii) The Conditions for the Technique's Optimal Performance

To enhance the method: (1) supplement degree analysis with contextual understanding of the specific criminal context, including the nature of activities, individual roles, and social factors (this avoids misinterpretations), (2) compare the analyzed network with baseline or reference networks created using random or hypothetical data (this helps identify influential criminals by establishing a benchmark for comparison), (3) integrate degree analysis with other network analysis techniques for a comprehensive understanding of the network's structure and dynamics, enhancing the insights gained.

### iii) Research Papers that have Employed the Technique

Bright et al. [67] examined the efficacy of five different law enforcement interventions in disrupting and dismantling criminal networks. The study focused on evaluating three measures at each step of the interventions, which included: (1) network degree centralization, (2) the number of active components within the network, and (3) the size of the largest component.

Agarwal et al., [68] collected a dataset from Twitter, which included 3.2 million distinct users and over 12 million tweets. The data was categorized into three awareness groups to identify influential individuals within each category. The researchers analyzed various dynamic features of each user, such as their in-degree, out-degree, favorite count, and activity on social media platforms (SMPs). These features were used to rank the users across the entire dataset. Bright and Delaney, [69] investigated how a criminal network's structure and function evolved over time. They computed centrality measures, specifically degree and betweenness, for each member of the network at different time intervals.

Colladon and Remondi [70] conducted research on the significance of social network metrics and introduced novel network mapping techniques that are not typically used in anti-money laundering practices. They analyzed variables such as in-degree, out-degree, and their combined all-degree measures. The researchers discovered that there are typically strong correlations between the in-degree, out-degree, and all-degree variables, indicating a normal relationship. Petersen et al. [71] introduced a knowledge management strategy for visualizing potential secondary effects following the removal of a node, enabling investigators to explore "what if" scenarios in criminal network analysis.

**Table 3:** Evaluating papers that have employed Degree-Based Analysis

| Paper/ Year | Dataset | Scalability | Interpretability | Accuracy | Efficiency |
|---|---|---|---|---|---|
| [67] 2017 | Simulated network | Acceptable | Good | Good | Acceptable |
| [68] 2021 | Twitter dataset | Fair | Fair | Acceptable | Acceptable |
| [69] 2013 | Author collected | Fair | Unsatisfactory | Acceptable | Fair |
| [70] 2017 | Tacit Link Network | Unsatisfactory | Fair | Acceptable | Acceptable |
| [71] 2011 | N17 dataset | Fair | Acceptable | Good | Fair |

### 2.1.2 Shortest Path-Based Analysis Approach

#### a) Closeness Centrality-Based Model

Closeness centrality is a pivotal measure in network analysis, crucial for identifying influential nodes within a network by evaluating their proximity and accessibility to other nodes. This metric is particularly instrumental in the analysis of criminal networks, offering a quantifiable means to pinpoint influential criminals based on their connectivity level and potential to govern information flow. It calculates how close a node is to all other nodes, considering the shortest paths between them, thus providing insights into the node's centrality and influence.

In the arena of criminal social networks, closeness centrality acquires a nuanced significance. It enables the identification of crime leaders or central figures who hold substantial control or influence over the network's dynamics. These individuals are characterized by their shorter path lengths to all other members, signifying their capacity to efficiently coordinate criminal activities, distribute resources, or relay critical information across the network. The measure is computed as the inverse of the sum of the shortest path distances from a given node to all other nodes in the network, mathematically represented as $C(v) = 1 / \Sigma\, d(v, u)$, where $C(v)$ denotes the closeness centrality of node "$v$", $d(v, u)$ represents the geodesic distance between nodes "$v$" and "$u$", and $\Sigma$ symbolizes the summation over all other nodes. Fig. 4 shows the procedure of the closeness centrality.

The application of algorithms like Dijkstra's or Floyd-Warshall enhances the precision of assessing each criminal's closeness centrality, distinguishing those with higher values as more influential within the network. Such individuals are not only key players due to their access to critical information, resources, or connections but also serve as vital intermediaries or decision-makers. Their elevated centrality positions them as crucial nodes for the facilitation of illicit activities, enabling them to exert considerable control over other members and significantly impact the network's functionality and resilience.

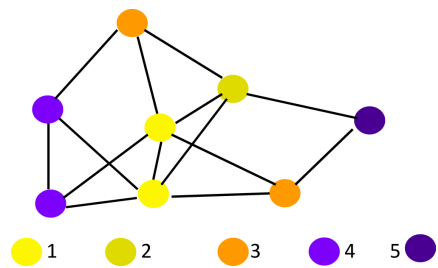

**Fig. 5:** The procedure of closeness centrality. Each node's legend number indicates the centrality rank of the node.

*i) The Rationale Behind the Usage of the Technique*

(1) <u>Quick Access to Information</u>: Leaders in criminal networks use their position to quickly spread instructions or collect information across the network. High closeness centrality indicates short paths to all nodes, showcasing their capability for efficient communication, crucial for coordinating and controlling network activities, (2) <u>Strategic Positioning for Control</u>: Those with high closeness centrality are well-placed to effectively influence the network. Their central position is vital for managing operations and upholding the criminal enterprise's integrity, and (3) <u>Uncovering Hidden Figures</u>: Criminal leaders often avoid detection by not being overtly connected or actively involved in criminal acts. Closeness centrality reveals these covert leaders through their efficient network communication, rather than visible connections.

*ii) The Conditions for the Technique's Optimal Performance*

The technique can be enhanced by: (1) a connected network is crucial (fragmented or isolated nodes/groups hinder closeness centrality's accuracy in assessing overall criminal influence), (2) relying solely on closeness centrality may not sufficiently identify influential criminals, necessitating consideration of specific criminal activities and roles, (3) efficient communication channels among criminals are vital for using closeness centrality to identify influential individuals (barriers undermine the centrality's ability to capture influence), (4) depending on network's characteristics, an appropriate centrality threshold is needed.

*iii) Research Papers that have Employed the Technique*

Calderon [26] examined network analysis to identify leaders within a large mafia network. The study focused on data gathered from an extensive investigation of the 'Ndrangheta, a Calabria-based mafia organization in Southern Italy. Operation Infinito successfully uncovered multiple mafia families and monitored their operational meetings. The author employed various metrics, including closeness centrality, degree centrality, and betweenness centrality, in the analysis. Shafia and Chachoo [27] studied the impact of social media platforms, particularly Facebook, on the spread of criminal propaganda. They formed smaller subnetworks to identify individuals and their connections, which required more in-depth examination. They utilized closeness centrality as a key concept. Yang [1] used SNA multiple times to extract crime networks and identify influential individuals within criminal organizations. The Fisher Discriminant Analysis Method was applied to determine a threshold for categorizing nodes into distinct groups. SNA facilitated crime network mining and the identification of key figures within the network. Memon [28] highlighted the importance of strong connections in evaluating node centrality. The author employed metrics like closeness and betweenness centrality, focusing on identifying the shortest paths and their lengths between nodes. This approach recognized the significance of strong ties.

**Table 4:** Evaluating papers that have employed Closeness Centrality

| Paper/Year | Dataset | Scalability | Interpretability | Accuracy | Efficiency |
|---|---|---|---|---|---|
| [26] 2014 | Chalonero dataset | Fair | Good | Good | Fair |
| [28] 2012 | Krebs's 9/11 dataset | Fair | Acceptable | Acceptable | Good |

**b) Betweeness Centrality-Based Model**

Betweenness centrality is a crucial metric for identifying influential individuals within networks, particularly in the context of criminal social networks. This measure quantifies the degree to which a node (representing an individual within the network) acts as a bridge along the shortest paths between other nodes, effectively serving as an intermediary that connects various segments of the network. By calculating the number of shortest paths passing through a specific node, betweenness centrality highlights nodes with significant influence over the network's structure and the flow of information within it. It is defined and computed as follows:

$$C(v) = \sum_{n \neq v \neq m \notin V} \frac{\sigma_{n,m}(v)}{\sigma_{nm}}$$

- $V$ is the set of nodes in the graph,
- $\sigma_{n,m}$ is the total number of shortest paths from node $n$ to node $m$.
- $\sigma_{n,m}(v)$ is the number of those paths that pass through $v$.

To compute betweenness centrality accurately, algorithms such as Brandes or the Girvan-Newman are utilized. These algorithms determine the shortest paths between every pair of nodes in the network and calculate the proportion of these paths that include each node. Nodes scoring high in betweenness centrality are recognized as critical connectors. They not only link different parts of the network but also control the dissemination of information, thereby holding potential power and influence within the network. By applying these methodologies, nodes with the highest betweenness centrality scores are identified as having substantial influence within the social network. Such nodes often play pivotal roles as crime leaders or key figures in criminal networks, as they manage and facilitate the flow of communication and resources.

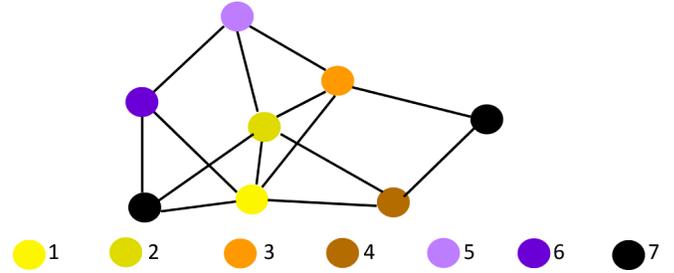

**Fig. 6:** The procedure of betweenness centrality. Each node's legend number indicates the centrality rank of the node.

*i) The Rationale Behind the Usage of the Technique*

In criminal networks, interconnected subgroups or communities are common. Betweenness centrality helps identify individuals who connect or mediate between these groups. In such networks, effective information dissemination is crucial for operational success. Individuals with high betweenness centrality regulate information flow and influence decision-making. Their removal or arrest can fragment the network. High betweenness centrality is attributed to individuals who appear frequently on the shortest paths within the criminal network, indicating their crucial roles within it. This metric holds relevance in criminal investigations, empowering analysts to prioritize their attention to individuals who exert influence in the network.

*ii) The Conditions for the Technique's Optimal Performance*

(1) <u>Dynamic vs. Static Analysis</u>: Criminal networks can be dynamic, with relationships and nodes changing over time. Static analysis might not accurately reflect the current structure or identify current leaders. An optimal approach would involve dynamic analysis, considering the evolution of the network over time to identify leaders who may change as the network evolves, (2) <u>Weighted vs. Unweighted Edges</u>: Incorporating the strength or weight of the connections (e.g., frequency of interactions, importance of transactions) can enhance the detection of leaders. In weighted networks, betweenness centrality calculations that consider these weights can provide a more nuanced understanding of influence and control within the network, and (3) <u>Combination with Other Measures</u>: Combining betweenness centrality with other network centrality measures (such as closeness centrality, degree centrality, and eigenvector centrality) can provide a more comprehensive view of an individual's role and influence in a criminal network. This holistic approach can improve the accuracy of identifying leaders.

*iii) Research Papers that have Employed the Technique*

Malm et al. [29] analyzed co-offending in various criminal enterprise groups. They used data from multiple police systems and employed betweenness centrality to study the structure and composition. The study revealed distinct co-offending patterns across different crime groups. Décary-Hétu and Dupont [30] evaluated the effectiveness of SNA in enhancing information about cybercriminals and identifying subjects for further investigation. They demonstrated that SNA, including betweenness centrality and other measures, provides scientific and unbiased metrics for identifying key actors.

Fidalgo et al. [31] conducted a study to explore the identification of potentially significant nodes in fraud networks. Their approach involved assessing the relationship between control and influence by using betweenness centrality to measure bridging centrality. The bridging centrality metric of a node is derived from its betweenness centrality and bridging coefficient. Taha and Yoo [32] introduced a forensic analysis approach to identify influential criminals in a criminal network using edge betweenness centrality. Their method calculates the shortest-path edge betweenness for each edge. A Minimum Spanning Tree (MST) is constructed for the network based on these weights. Each node, denoted as u, is assigned a score representing the number of nodes in the MST that rely on u for their existence. Influential nodes are identified by ranking them according to their score.

Calderoni and Superchi [33] conducted a study that investigated the characteristics of criminal leadership by examining the involvement of leaders in meeting and telephone communications. They also compared the meeting and wiretap networks to identify leaders. The findings revealed a significant correlation between high betweenness centrality and the likelihood of being a criminal leader. Taha and Yoo [34] proposed a forensic analysis method to detect influential criminals in a network. The method focuses on critical communication pathways and involves calculating betweenness centralities of nodes to estimate their impact on information flow. The method also considers the betweenness centralities of nodes connected to the path, providing a comprehensive evaluation of path importance.

Grassi et al. [35] studied betweenness centrality to identify criminal leaders in a meeting participation network. Despite expected correlations, different forms of betweenness centrality yielded distinct rankings for nodes. Dual projection methods were generally more effective than traditional approaches in identifying criminal leaders.

**Table 5:** Evaluating papers that employed the Betweenness Centrality.

| Paper/Year | Dataset | Scalability | Interpretability | Accuracy | Efficiency |
|---|---|---|---|---|---|
| [31] 2022 | Telecom dataset | Fair | Good | Acceptable | Unsatisfactory |
| [32] 2017 | Krebs's 9/11, Enron email corpus | Unsatisfactory | Good | Good | Acceptable |
| [33] 2019 | Aemilia meeting, Minotauro meeting | Unsatisfactory | Acceptable | Acceptable | Fair |
| [34] 2019 | Caviar, Enron email corpus | Fair | Good | Good | Fair |
| [35] 2019 | Operazione Infinito | Unsatisfactory | Fair | Good | Good |

## 2.2 Local Analysis Scope

### 2.2.1 Node Similarity-Based Model

In the analysis of criminal networks, the concept of node similarity centrality plays a pivotal role in detecting crime leaders and understanding the intricate connections within such networks. This metric illuminates the extent to which a particular node—representing an individual within the criminal network—shares connections with other influential nodes. Essentially, it gauges the level of interconnectedness of a criminal with other high-profile criminals based on shared neighbors, offering insights into the hierarchical structure and potential influence within the network.

To effectively identify these key individuals, it is crucial to employ a similarity metric that captures the characteristics or attributes that render a criminal influential. This involves considering factors like the nature and gravity of criminal activities, in addition to the breadth and significance of their connections. Various node similarity metrics are instrumental in this process, including but not limited to Jaccard Similarity, Adamic-Adar Similarity, Preferential Attachment, Cosine Similarity, Pearson Correlation Coefficient, and Euclidean Distance. These metrics offer diverse lenses through which to view and evaluate the connections and similarities between nodes.

Moreover, the structural similarity, determined by the connections between nodes, complements the attribute similarity measures to provide a holistic view of an individual's role within the network. High centrality scores, derived from network centrality measures, spotlight nodes with significant influence due to their strategic positions in the network. This structural analysis, when combined with an examination of individual characteristics—including demographics, criminal history, and affiliations—through similarity metrics like cosine or Jaccard similarity, enriches the understanding of how individuals are linked and how influential figures emerge within criminal networks.

By integrating both structural and attribute similarity measures, the approach to identifying crime leaders becomes significantly more comprehensive. This fusion enables a deeper dive into the complex fabric of criminal networks, ensuring a robust framework for detecting and understanding the nuances of criminal influence and connectivity.

*i) The Rationale Behind the Usage of the Technique*

Structural equivalence suggests that nodes with similar connections have similar roles in the network. Individuals who are structurally equivalent to influential criminals may hold influential positions in the criminal network. This helps to identify nodes resembling influential criminals based on their connections. By systematically examining connections and relationships, we can identify individuals who exhibit similar behavioral patterns and characteristics to those of known influential criminals. Within the network of connections that we evaluate, certain nodes emerge as potential influential criminals. These individuals exhibit a level of influence that is comparable to known influential criminals within the criminal activities.

*ii) The Conditions for the Technique's Optimal Performance*

The criminal network representation should accurately capture interactions and relationships while reflecting its specific context and dynamics. Choosing an appropriate similarity metric is vital, as different metrics yield different results based on the network structure and criminal connections. A threshold may be needed to determine significant connections, enhancing accuracy in identifying influential criminals. Comparing node similarity centrality scores against relevant baselines is crucial. Validating results with additional information and considering temporal dependencies can enhance our understanding of network influence, leading to a more precise analysis. We can uncover the underlying mechanisms hat drive network influence over time.

*iii) Research Papers that have Employed the Technique*

Berlusconi et al. [36] contended that intelligence and investigation activities could suffer adverse consequences as law enforcement agencies might overlook certain individuals and connections. They showcased how node similarity can detect potential missing links within criminal networks, even when the available information is inherently noisy or incomplete. Tundis et al. [37] utilized node similarity to examine the resemblances and connections between criminals involved in Organized Crime and Terrorist Networks. Their objective was to uncover both similarities and clusters of users associated with illegal activities, such as drugs, weapons, and human trafficking. Calderoni et al. [38] utilized community analysis techniques to investigate the arrangement of a criminal network, which depicted the extent of individuals' joint participation in meetings. They employed a node similarity measure, operating under the assumption that nodes exhibiting higher similarity are more likely to share the same label. Kumari et al. [39] presented two intelligent techniques for deceiving community detection algorithms with the aim of concealing nodes within a network. They employed node-based matrices, persistence scores, and safeness scores to define optimization problems that would confuse the CDAs.

**Table 6:** Evaluating papers that employed Node Similarity centrality.

| Paper/Year | Dataset | Scalability | Interpretability | Accuracy | Efficiency |
|---|---|---|---|---|---|
| [36] 2016 | Oversize dataset | Acceptable | Good | Good | Acceptable |
| [37] 2019 | AboutIsis dataset | Unsatisfactory | Fair | Acceptable | Fair |
| [39] 2021 | Domestic terror network, Dolphins datasets | Unsatisfactory | Fair | Good | Good |

### 2.2.2 Local Clustering Coefficient-Based Model

In the realm of criminal networks, the local clustering coefficient emerges as a pivotal measure within network analysis, aimed at delineating the degree of interconnectedness or clustering among nodes in a network. This metric proves particularly instrumental in identifying influential criminals or individuals wielding significant influence within their immediate circles. The essence of the local clustering coefficient lies in its ability to quantify the extent of closeness among the neighbors of a given node, denoted as $C_i$, which essentially reflects how interconnected a node's immediate network is. It's computed by comparing the actual number of links present among neighboring nodes to the maximum potential links that could exist within that locale, employing algorithms like the Watts-Strogatz algorithm or triangle counting for precise calculation. For undirected node, the local clustering coefficient $C_i$ for a node $i$ is defined as follows: $C_i = 2 E_i / (k_i (k_i - 1))$, where $E_i$ is the number of connections (edges) existing between the neighbors of node $i$, $k_i$ is the degree of node $I$ (i.e., the number of neighbors of node $i$), and the factor of 2 in the numerator is used to adjust for the fact that each connection between two neighbors is counted twice in undirected graph (one for each direction).

The application of this measure within criminal networks is strategic, enabling the calculation of the local clustering coefficient for each individual within the network. This process sheds light on the intricate web of relationships and the degree of interconnectedness among criminals, pinpointing those with notably high local clustering coefficients. Such individuals are characterized by robust connections with their neighbors, suggesting potential clusters of criminal activity.

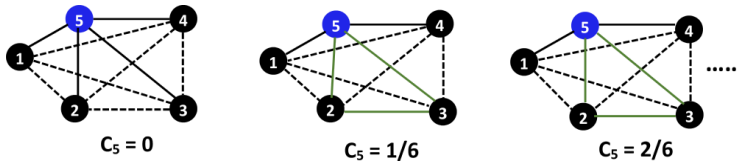

**Fig. 7:** The procedure of local clustering coefficient is illustrated in the figure.

*i) The Rationale Behind the Usage of the Technique*

The local clustering coefficient identifies criminal clusters with strong interconnections. These clusters foster cohesive relationships and collaboration among criminals, providing insights into influential individuals. High clustering coefficients indicate close connections between specific criminals, suggesting collaboration in criminal activities. Also, the coefficient highlights bridge nodes that connect different clusters, playing crucial roles in linking isolated parts of the network. These bridge nodes act as critical intermediaries, facilitating the exchange of illicit goods, services, or knowledge across different criminal groups or activities. They possess a unique position within the network.

*ii) The Conditions for the Technique's Optimal Performance*

To enhance the technique: (1) identify nodes with high local clustering coefficients, indicating potential influence due to their highly interconnected neighbors, (2) consider nodes with high degree centrality, suggesting influential criminals with extensive connections, (3) pay attention to nodes connecting multiple high-degree nodes, acting as bridges between distinct clusters or groups, enabling influence across the criminal network, and (4) analyze the clustering patterns of a node's neighbors to identify tightly knit criminal subnetworks where the node's presence signifies influence within its immediate vicinity, (5) the network should have a more uniform distribution of connections across nodes.

*iii) Research Papers that have Employed the Technique*

Agreste et al. [16] examined the network structure of a Mafia organization, documenting its development over time and emphasizing its adaptability to interventions aimed at membership targeting. The researchers proposed a two-stage approach, where the criminal network was initially divided into subgroups using a clustering algorithm. They then calculated the Average Clustering Coefficient for each vertex in relation to its degree and consistently found it to be greater than 0.6. Ozgul and Erdem [40] introduced a measure of resilience for criminal networks, which they applied to two actual criminal networks. They examined the resilience outcomes in relation to various factors. To assess the resilience, the authors utilized the Average centrality of leaders and the Clustering coefficient.

Catanese et al. [41] introduced LogAnalysis, a forensic system designed to assist forensic investigators in comprehending the hierarchical structures within criminal organizations. LogAnalysis offers the capability to calculate both the global clustering coefficient of a given phone call network and the local clustering coefficient of nodes. Song et al. [42] examined two distinctive features of small-world networks: the local clustering coefficient and the global characteristic path length. Their findings revealed that individuals within the fake review group exhibited lower friend counts and were more inclined to provide negative ratings, particularly with ratings of 1 or 2.

**Table 7:** Evaluating papers that employed Local Clustering Coefficient

| Paper/Year | Dataset | Scalability | Interpretability | Accuracy | Efficiency |
|---|---|---|---|---|---|
| [16] 2016 | Sicily Mafia gangs datasets | Fair | Good | Acceptable | Acceptable |
| [40] 2015 | Operation Cash, Ex-inmates Network | Unsatisfactory | Unsatisfactory | Fair | Acceptable |
| [41] 2013 | Phone call networks | Acceptable | Fair | Good | Good |
| [42] 2017 | Yelp online review | Unsatisfactory | Unsatisfactory | Fair | Acceptable |

## 3. Clustering-Based Analysis

### 3.1 Spatial-Based Analysis

#### 3.1.1 Global-Based Analysis

**a) Hierarchical-Based Clustering**

Hierarchical-based clustering is a powerful tool for identifying influential individuals within criminal social networks, especially when considering the challenge of detecting crime leaders. This method assesses the likeness between nodes in the network, employing both agglomerative and divisive approaches to create a hierarchical structure of clusters that reflect the relationships and similarities among individuals.

In the agglomerative approach, each node starts as a separate cluster, and similar clusters merge based on a linkage criterion. This process continues until a single cluster represents the entire network. Conversely, divisive clustering begins with the network as one cohesive cluster, which then gradually splits based on dissimilarity thresholds. This process continues until it reaches the level of individual nodes. The position of nodes within this hierarchy reveals their influence within the network, identifying those who hold pivotal roles in the criminal organization.

The technique adds another layer of analysis by considering the geographical aspects of the criminal network. This approach enables the identification of spatially cohesive clusters of influential criminals, adding a critical dimension to understanding the structure and operation of criminal networks. By successively merging or splitting clusters at different levels, spatial global-based hierarchical clustering creates a detailed map of how criminal influence and operations are distributed geographically, offering insights into the influential figures in the networks

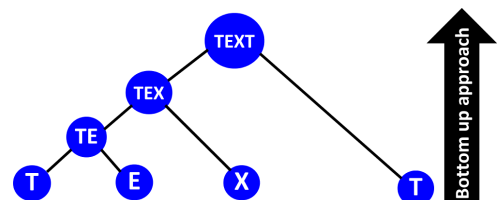

**Fig. 8:** The general procedure of the hierarchical-based clustering approach is illustrated in the figure.

*i) The Rationale Behind the Usage of the Technique*

Hierarchical clustering identifies patterns in a criminal network by grouping similar individuals, revealing clusters that represent criminal organizations, factions, or hierarchies. This method captures hierarchical structures where influential individuals hold higher positions. Organizing clusters into a tree-like structure helps identify key players at different influence levels. Through the analysis of connections between clusters and the identification of individuals bridging different parts of the network, hierarchical clustering aids in pinpointing central figures.

*ii) The Conditions for the Technique's Optimal Performance*

The technique is effective when the network has moderate to high connection density, but sparse networks present challenges in identifying influential criminals due to limited information. Choosing an appropriate similarity or distance measure (e.g., Euclidean distance or Jaccard index) is crucial. Linkage methods (e.g., single linkage) yield varying results, so selecting the right method aligns with the network's characteristics is crucial. Defining the optimal number of clusters requires a clear criterion.

*iii) Research Papers that have Employed the Technique*

A forensic system called CrimeNet Explorer, developed by Xu and Chen [43], aids in uncovering criminal networks and their leaders. It utilizes techniques like the concept space approach, hierarchical clustering, social network analysis, and multidimensional scaling. Hierarchical clustering divides the network into subgroups based on relationship strength, facilitating the identification of significant criminals. Kazmi et al. [44] compared ten methods to detect roles and influential leaders in terrorist networks, finding hierarchical-based clustering to be the most effective. Afra and Alhajj [45] used crime incident reports to construct a criminal graph, connecting criminals based on co-occurrence in the reports. They applied hierarchical clustering to uncover influential criminals, reveal hidden relationships, and identify hierarchical criminal groups.

**Table 8:** Evaluating papers that employed Hierarchical-based clustering.

| Paper/Year | Dataset | Scalability | Interpretability | Accuracy | Efficiency |
|---|---|---|---|---|---|
| [43] 2005 | Narcotic and gang networks | Unsatisfactory | Good | Good | Acceptable |
| [45] 2021 | Author collected | Unsatisfactory | Good | Fair | Fair |

**b) Spatial Random Graph Distribution-Based Clustering**

Spatial Random Graph Distribution-Based Clustering combines spatial data with network analysis to uncover clusters within criminal social networks, aiming to identify influential criminals or key players. This method acknowledges the significance of geographical proximity by assigning spatial coordinates to each network node, reflecting their real-world locations. By doing so, it emphasizes the spatial relationships among individuals, enhancing the understanding of how these networks operate within specific geographical areas.

The algorithm employs techniques like k-means or DBSCAN, integrating both spatial and network proximity to organize nodes into clusters. This approach not only considers the network connections among individuals but also leverages spatial distances to paint a comprehensive picture of the network's structure. The model is particularly adept at detecting clusters of influential criminals, assuming these individuals are embedded in tightly-knit groups characterized by both strong network ties and significant spatial clustering.

*i) The Rationale Behind the Usage of the Technique*

Criminal activities often show spatial patterns, with criminals operating in specific regions. This approach considers spatial relationships to detect clusters of criminals in proximity, indicating hotspots. Criminal networks have complex structures with various relationships like co-offending and communication. This method analyzes connections between individuals to identify tightly connected subgroups, revealing cohesive criminal groups. Influential criminals exhibit distinct behaviors, and this approach can detect anomalies within the network.

*ii) The Conditions for the Technique's Optimal Performance*

To enhance the method: (1) incorporate spatial attributes into the criminal social network to capture accurate spatial patterns and connections, (2) collect sample data to ensure meaningful results, (3) account for spatial relationships of criminals accurately, as the method assumes a decrease in interaction probability with increasing spatial distance, (4) carefully select parameter values (e.g., spatial decay function) based on the network characteristics, (5) select appropriate statistical methods (e.g., maximum likelihood estimation).

*iii) Research Papers that have Employed the Technique*

Agarwal and Toshniwa [46] identified influential local leaders in hazard networks, including criminal networks. Their algorithm assigned ranking scores to weakly connected neighbors, considering random spatial information. Factors like outer degree, inner degree neighbors, epsilon, and damping factors strongly influenced the ranking relationship. Ficara et al. [23] conducted simulations to disrupt Mafia networks, using centrality metrics (degree, betweenness, closeness) to identify influential criminals. The intervention strategy randomly eliminated one actor at each step, assessing network integrity with three measures. Duijn et al. [47] studied resilience in criminal networks when disrupted. They found that targeting influential leaders weakens the network. Three recovery mechanisms were used to simulate network resilience.

**Table 9:** Evaluating papers that employed Random Graph Distribution

| Paper/Year | Dataset | Scalability | Interpretability | Accuracy | Efficiency |
|---|---|---|---|---|---|
| [46] 2020 | Natural hazard dataset | Unsatisfactory | Acceptable | Acceptable | Good |
| [23] 2022 | Montagna operation | Fair | Good | Acceptable | Good |
| [47] 2014 | Simulated crime dataset | Unsatisfactory | Good | Fair | Acceptable |

### 3.1.2 Local-Based Analysis

**a) Spatial Diffusion-Based Clustering**

Local Spatial Diffusion Clustering is an advanced technique specifically designed for identifying influential leaders within criminal networks by examining both their social connections and spatial proximity. This method intricately combines the analysis of spatial diffusion patterns—how criminal behavior spreads through a network—and the underlying social relationships among individuals to effectively detect clusters of criminal activity. By simulating the dissemination of criminal behaviors among interconnected individuals, it considers factors like past criminal associations, leveraging these insights to influence the diffusion process.

The technique employs metrics to quantify an individual's influence within the network, integrating these measurements with density-based clustering results and spatial constraints. This ensures a more precise identification of clusters that are not only based on social ties but also consider the physical closeness of individuals, recognizing that both elements are crucial in the propagation of criminal activities. Through the propagation of cluster labels, influenced by the calculated diffusion metrics and spatial clustering methods, Local Spatial Diffusion Clustering is adept at pinpointing those criminals who play pivotal roles in orchestrating crimes within their networks.

*i) The Rationale Behind the Usage of the Technique*

Spatial diffusion clustering focuses on local interactions/connections, considering geographical or network proximity. This helps understand how criminal behavior disseminates, especially in networks where physical location and proximity are crucial. By analyzing diffusion patterns and network centrality, influential individuals can be identified. This helps identify "hotspots" of criminal activities, offering insights into the Key criminal hubs can be detected and the direction of criminal flow can be discerned.

*ii) The Conditions for the Technique's Optimal Performance*

To enhance the method: (1) ensure the network has clear community structure, with densely connected criminals sharing similar characteristics or roles within communities and sparser connections between communities, (2) employ various models (e.g., threshold model, cascade model) based on the characteristics of the network, (3) select seed nodes that cover different communities, (4) carefully tune parameters, such as the propagation threshold and activation probability (likelihood of adopting influence).

*iii) Research Papers that have Employed the Technique*

Meneghini et al. [48] introduced a methodological approach for estimating a criminal trafficking network across different local spatial geographical levels. This methodology focuses on identifying the most probable routes within the criminal network that can be targeted and exploited by criminals.

Taha and Yoo [49] introduced CLDRI, a forensic analysis system to identify key individuals in a criminal organization. CLDRI quantifies their influence by considering factors like local spatial relationships and information diffusion among connected nodes.

**Table 10:** Evaluating papers employed Spatial Diffusion Clustering

| Paper/Year | Dataset | Scalability | Interpretability | Accuracy | Efficiency |
|---|---|---|---|---|---|
| [48] 2020 | Illicit goods trafficking dataset | Fair | Fair | Good | Good |
| [49] 2015 | Enron email corpus | Unsatisfactory | Fair | Good | Fair |

**b) Local Density-Based Clustering (LSBC)**

LDBC stands out as a strategic approach tailored for unveiling influential figures within the spheres of criminal networks. This methodology leverages the concept of local density to efficiently cluster individuals, pinpointing those wielding substantial sway within these illicit circles. Central to LDBC's operation is the meticulous analysis of the connections' density encircling each node within the network, thereby facilitating the identification of influential criminal clusters. The process initiates with the selection of an arbitrary core point, subsequently extending through the identification of all points that are density-reachable from this core. This expansion is recursive, fostering the connection of core points and augmenting the clusters progressively until the reachability threshold is met. Such clusters, characterized by a dense concentration of influential criminals, are indicative of significant criminal cohorts.

Fundamental to LDBC's clustering mechanics are two critical parameters: epsilon (ε) and minPts. Epsilon denotes the radius that defines the neighborhood's reach around a point, while minPts specifies the minimum number of points required to constitute a dense region. A data point is considered density-reachable and thus part of a cluster if it lies within the epsilon vicinity of a core point, provided that the latter is a core point. This framework also serves as the cornerstone for the expansion and refinement of clusters. Points that do not satisfy these criteria are deemed outliers, underscoring their lack of integration within the dense regions.

*i) The Rationale Behind the Usage of the Technique*

Density-based clustering, like DBSCAN, is adept at handling non-linear and irregular cluster shapes, enabling the detection of intricate patterns in networks. This makes it suitable for identifying influential criminals in cohesive groups. It can handle noise and outliers by treating them as separate, allowing focus on dense clusters likely to contain influential criminals. It enables analyzing large criminal social networks, making it practical to identify influential criminals in real-world scenarios.

*ii) The Conditions for the Technique's Optimal Performance*

To enhance the method: (1) carefully select parameters for the clustering algorithm, (2) select the distance metric based on the data characteristics (e.g., using Jaccard similarity if the criminal network is represented as a graph, and Euclidean distance if the data is in vector form), (3) decide whether to include outliers in the clustering results or treat them separately (4) incorporate domain knowledge and expertise of the criminal social network to significantly enhance the clustering performance.

*iii) Research Papers that have Employed the Technique*

Everton et al. [50] analyzed the evolution of the Noordin Top terrorist network and its changing structure over time. They investigated how the network's goals and strategies employed by authorities affected its interconnectedness and local spatial density. The authors observed that some networks tend to become more internally dense and centralized as time goes on. Gunnell et al. [51] proposed a method to identify individuals associated with gangs, comprehend gang activities, and gain insights into gang structure and organization. They found that the "gang links" sub-network had the highest density, indicating higher levels of involvement.

**Table 11:** Evaluating papers that employed Density-Based Clustering

| Paper/Year | Dataset | Scalability | Interpretability | Accuracy | Efficiency |
|---|---|---|---|---|---|
| [50] 2015 | Noordin Top network | Acceptable | Good | Fair | Good |
| [51] 2016 | Greater Manchester police data | Fair | Acceptable | Acceptable | Good |

**c) Local Fuzzy C-Means (FCM) Clustering**

The Local Fuzzy C-Means Clustering (FCM) approach is adept at identifying influential criminals within a network by leveraging a fuzzy clustering algorithm that focuses on the local characteristics and connectivity among individuals in the criminal network. This method assigns membership degrees to each criminal, reflecting their level of influence and association within the network. Higher membership degrees denote a stronger influence, acknowledging that individuals can belong to multiple clusters, each with varying degrees of influence.

The algorithm calculates cluster centroids as weighted averages based on data points and their corresponding membership degrees, representing the core points and defining characteristics of each cluster. FCM iteratively fine-tunes these membership degrees and centroids to achieve an optimal clustering solution. The introduction of the fuzziness parameter (m) allows for controlling the ambiguity of assignments, where higher values permit more flexible associations across multiple clusters. This parameter is crucial for handling data with overlapping or uncertain boundaries between clusters, enabling a more nuanced identification of influence within the criminal network. By focusing on both the connections among criminals and the flexibility in cluster membership, the Local Fuzzy C-Means Clustering approach provides a sophisticated tool for detecting crime leaders and understanding their roles within criminal networks.

*i) The Rationale Behind the Usage of the Technique*

Criminals can have different impacts based on connections, criminal activities, and social reputation. FCM clustering assigns membership degrees to capture this variability in influence. In a criminal network, some criminals can have varying degrees of influence in multiple clusters. FCM allows criminals to belong to multiple clusters, reflecting their influence in different criminal activities. This helps identify influential criminals spanning across various groups, crucial for understanding their overall impact in a network. FCM is robust against noise and outliers as it considers membership degrees rather than relying solely on distance measures. FCM provides a valuable approach for clustering tasks that involve uncertainty and overlapping clusters.

*ii) The Conditions for the Technique's Optimal Performance*

The method can be enhanced by: (1) select an appropriate membership function to match data characteristics, with options like Gaussian, exponential, and sigmoid functions, (2) choose a reasonable number of clusters based on prior knowledge or techniques like elbow method or silhouette analysis, (3) select network features that capture factors like connections, propagation, and social attributes play a role, (4) set convergence criteria (e.g., maximum iterations, membership value threshold) carefully (too few iterations yield suboptimal results, while excessive iterations increase computation time without significant clustering improvement), and (5) cluster validity indices to help assess clustering quality.

*iii) Research Papers that have Employed the Technique*

Premasundari and Yamini [52] utilized a local Fuzzy C-Means approach to perform clustering on crime rates. They developed a novel multiple clustering model and assessed its effectiveness using the USArrests dataset. The outcomes were then employed to forecast the likelihood of crime occurrence by visually analyzing the crime patterns across different states in the United States. Sivanagaleela and Rajesh [5] introduced a fuzzy C-Means algorithm, which was suggested as a suitable method for clustering crime data related to various cognizable crimes, Murder, Theft, Burglary, and Robbery. By employing the fuzzy clustering technique, this algorithm effectively identifies areas with higher crime rates. conducted a study on factor clustering analysis concerning violent crimes.

**Table 12:** Evaluating papers that employed Local Fuzzy C-Means

| Paper/Year | Dataset | Scalability | Interpretability | Accuracy | Efficiency |
|---|---|---|---|---|---|
| [52] 2019 | USArrests dataset | Fair | Unsatisfactory | Fair | Acceptable |
| [5] 2016 | Indian crime data | Unsatisfactory | Unsatisfactory | Fair | Fair |

## 3.2 Spatiotemporal -Based Analysis

### 3.2.1 Spatiotemporal Diffusion-Based Analysis

Spatiotemporal diffusion-based clustering integrates spatial and temporal data to uncover patterns within networks, especially effective in identifying influential figures within criminal networks. This method leverages diffusion models, such as the Hawkes process, to analyze how criminal activities proliferate across both locations and time, highlighting the propagation patterns of criminal behaviors and the roles individuals play within these networks.

By combining common clustering algorithms, including k-means, DBSCAN, and spectral clustering, with network centrality measures like degree and betweenness, this approach effectively partitions data points into clusters. These clusters are based on their diffusion characteristics and quantify an individual's influence within the criminal network. This dual analysis not only identifies the spread of criminal activities but also pinpoints individuals with significant influence scores, indicating their critical roles in the diffusion process.

Furthermore, analyzing spatiotemporal patterns aids in detecting hotspots and temporal trends of criminal behavior, offering insights into the dynamics of crime propagation. The methodology incorporates comprehensive spatial and temporal data—geographic coordinates, timestamps, and diffusion process attributes—to investigate the interactions and interdependencies between the spatial and temporal dimensions of crime networks.

*i) The Rationale Behind the Usage of the Technique*

Spatiotemporal diffusion analysis can unveil patterns in the spread of criminal activity, offering valuable insights into the structure, organization, and dynamics of criminal networks. By studying historical data on criminal behavior, this analysis can help identify potential future hotspots or areas where criminal activities are likely to emerge or increase. Criminal networks are complex systems with intricate relationships and dynamics. Understanding spatiotemporal diffusion enables us to grasp how criminal behaviors propagate through these networks, including their speed and direction. This comprehension is crucial for developing effective strategies to disrupt or dismantle the network. Overall, Spatiotemporal Diffusion Analysis is a valuable tool for understanding, predicting, and managing the spread of phenomena across space and time.

*ii) The Conditions for the Technique's Optimal Performance*

Enhancements can be achieved by: (1) ensuring accurate network representation for identifying influential criminals, (2) refining spatial and temporal resolution to capture the geographical and temporal context of criminal activities, (3) selecting an appropriate propagation model (e.g., epidemic models, influence maximization models, or information cascades models) to accurately simulate diffusion within the criminal network, and (4) employing techniques like network centrality (e.g., degree centrality) and diffusion-based measures (e.g., influence scores, cascading probabilities) to quantify individual influence in the network. It is important to validate and evaluate the results of Spatiotemporal Diffusion Analysis. Comparison with ground truth data, external validation measures, or sensitivity analyses can help assess the accuracy.

*iii) Research Papers that have Employed the Technique*

Zhao and Tang et al. [53] provided a comprehensive overview of urban crime. Their study examined environmental and social criminal theories and utilized analysis techniques to gain insights from geospatial and temporal crime data. They also identified influential criminal leaders by studying criminal spatiotemporal diffusion patterns. Park and Tsang [54] proposed a framework to identify and visualize influential individuals in co-offending networks. They considered the temporal aspect by using centrality measures to detect key players in criminal networks. Their framework captures temporal changes and provides insights into network dynamics. Siriwat and Nijman [55] studied crime rates in Thailand to uncover spatial and temporal patterns. Their analysis aimed to identify and analyze patterns related to the origin, transit, and destination of crimes, as well as temporal patterns over time.

**Table 13:** Evaluating papers that employed Spatiotemporal Diffusion

| Paper/Year | Dataset | Scalability | Interpretability | Accuracy | Efficiency |
|---|---|---|---|---|---|
| [54] 2013 | Canadian RCMP dataset | Unsatisfactory | Good | Fair | Unsatisfactory |
| [55] 2023 | African Rivers dataset | Unsatisfactory | Acceptable | Acceptable | Fair |

### 3.2.2 Spatiotemporal Random Graph-Based Analysis

The spatiotemporal random graph-based models offer a sophisticated framework for understanding the intricacies of criminal networks by integrating both spatial and temporal dimensions of criminal activities. These models are constructed on a foundation that represents the social network of criminals, with nodes symbolizing individuals and edges depicting the relationships or interactions among them, all of which are enriched with spatiotemporal attributes. This mathematical depiction is critical in analyzing the connections between individuals involved in criminal behavior, providing a comprehensive view of how these activities are geographically and temporally interconnected.

By employing a variety of analytical techniques, such as network centrality measures (e.g., degree centrality, betweenness centrality), community detection algorithms, diffusion models, and statistical methods specifically tailored for spatiotemporal data analysis, the models facilitate a deeper understanding of the dynamics within criminal networks. This analysis identifies clusters of criminal activity that are not only interconnected but also evolve over space and time, offering valuable insights into the propagation patterns of criminal behaviors.

In the realm of detecting crime leaders within criminal networks, these spatiotemporal random graph-based models prove to be exceptionally valuable. By understanding the roles played by different individuals within these networks and identifying key actors based on their centrality and influence, it becomes possible to pinpoint influential criminals. This tailored approach aids in comprehending the hierarchical structures and dynamics of criminal activities.

*i) The Rationale Behind the Usage of the Technique*

The method enables us to comprehensively comprehend the structure of criminal networks by capturing both the spatial proximity and temporal dynamics of criminal interactions. By employing spatiotemporal random graph analysis, we gain insights into the evolution of the criminal social network, enabling us to detect emerging patterns like subgroup formation, the emergence of influential individuals, and changes in criminal information flow. Also, spatiotemporal random graph analysis can be utilized to develop predictive models for criminal behavior, enhancing proactive law enforcement measures.

*ii) The Conditions for the Technique's Optimal Performance*

Spatial and temporal resolutions must capture criminal interactions and network dynamics accurately. Enhanced resolutions aid in precise analysis, identifying patterns, subgroups, and influential individuals. Real-world data has uncertainties and missing information. Robust statistical methods, network algorithms, and imputation techniques address these issues and ensure reliable results. Metrics for identifying influential criminals should align with network objectives and characteristics, capturing various aspects of influence and power dynamics. Comparison with ground truth data, external validation measures, or sensitivity analyses can help assess the accuracy.

*iii) Research Papers that have Employed the Technique*

Berlusconi [56] introduced the spatiotemporal Random Graph Distribution to analyze temporal changes in structure. It combines qualitative analysis of wiretapped conversations with a quantitative element, using network statistics and exponential random graph models. This helps identify influential criminals. Win et al. [57] proposed an algorithm using fuzzy clustering and random initialization to identify potential criminal patterns in extensive spatiotemporal datasets. It detects patterns from large-scale datasets that encompass criminal activities. The study assessed crime rates for different locations. Griffiths et al. [58] studied the spatial and temporal behavior of UK-based Islamist terrorists. They analyzed the frequency and timing of their visits to different locations, aiming to assess if their movement patterns differed from the ordinary criminals.

**Table 14:** Evaluating papers that employed spatiotemporal random graph

| Paper/Year | Dataset | Scalability | Interpretability | Accuracy | Efficiency |
|---|---|---|---|---|---|
| [56] 2022 | Criminal collaboration network | Acceptable | Fair | Good | Acceptable |
| [57] 2019 | GTD dataset | Good | Acceptable | Good | Good |
| [58] 2017 | UK-based terrorist plots | Fair | Unsatisfactory | Acceptable | Acceptable |

## 4. Agent-Based Analysis

### 4.1 Eigen-Based Analysis

#### 4.1.1 PageRank-Based Analysis

PageRank is a graph algorithm tailored for identifying influential criminals within social crime networks. It functions by assigning importance scores to the network's nodes, considering the structure and connections within. Initially, every individual in the network is deemed equally important, embodying a uniform distribution of influence totaling a sum of 1. The underlying principle of PageRank posits that a node gains significance if it is linked to other substantial nodes. It is computed as follows:

$$PR(P_i) = \frac{1-d}{N} + d \sum_{P_j \in B_{P_i}} \frac{PR(P_j)}{L(P_j)}$$

where, $PR(P_i)$ is the PageRank of page $P_i$, $d$ is the damping factor, usually set to around 0.85, which represents the probability that a "random surfer" will continue clicking on links, $N$ is the total number of pages, and $L(P_i)$ is the number of outbound links on page $P_j$. In the realm of criminal networks, this translates to the notion that individuals connected to influential criminals are themselves more likely to wield influence.

The methodology of PageRank involves iterative recalculations of important scores, considering the connections between individuals and the magnitude of their influence. Each iteration allows an individual's importance to be disseminated among their connections, factoring in both the individual's significance and the strength of their connections. The process is sustained until the important scores stabilize, effectively pinpointing the network's key criminals. This stable state of scores reveals the leaders within criminal networks, leveraging the interconnected nature of their relationships to deduce their roles and influence.

*i) The Rationale Behind the Usage of the Technique*

The method considers the network structure to capture influence flow, identifying individuals with strong connections to other influential individuals. These well-connected individuals receive higher influence scores, suggesting that being linked to influential individuals enhances their own influence. This is crucial for identifying key players. PageRank provides an objective and quantitative measure of influence, enabling prioritization and ranking of individuals within the network. It is scalable and adept at handling large networks.

*ii) The Conditions for the Technique's Optimal Performance*

(1) assign appropriate weights to network edges to represent interactions between individuals and enhance analysis accuracy, (2) select the damping factor carefully based on the network's characteristics to ensure meaningful results, (3) set appropriate convergence threshold to balance computational efficiency and accuracy, considering the trade-off between smaller thresholds for higher precision, (4) implement strategies to handle sink nodes for unbiased results, (5) assign weights to nodes based on attributes like criminal records or affiliations.

*iii) Research Papers that have Employed the Technique*

Isah et al. [59] conducted a study to understand connections and community patterns in crime data, including conventional and cyber crimes, and predicting organized criminal networks. They used PageRank to analyze networks, identify influential nodes, detect subgroups, and assess network interconnectedness. Budur et al. [60] used probabilities generated by their model as weights for current edges to identify influential nodes. They calculated weighted PageRank scores from the weighted network and computed classical PageRank based on unweighted edges.

**Table 15:** Evaluating papers that have employed PageRank-based analysis

| Paper/Year | Dataset | Scalability | Interpretability | Accuracy | Efficiency |
|---|---|---|---|---|---|
| [59] 2015 | Darknet vendor-vendor network | Good | Unsatisfactory | Good | Acceptable |
| [60] 2015 | OFAC dataset | Fair | Fair | Good | Good |

#### 4.1.2 Eigenvector-Based Analysis

Eigenvector-based analysis stands as a potent method for pinpointing influential individuals within social networks, particularly within the realms of criminal networks. This approach is adept at identifying criminal leaders by leveraging the concept of centrality, which evaluates an individual's capacity to control or influence the flow of information or resources within the network. The essence of this technique lies in the creation of an adjacency matrix, where the relationships among criminals are encapsulated through the strength of their connections.

In analyzing criminal networks, eigenvector centrality plays a pivotal role. It not only assesses the direct connections of a node but also considers the influence of the nodes it is connected to. By doing so, it attributes higher centrality scores to nodes that maintain connections with other influential nodes, highlighting the key players who might be orchestrating criminal activities. The calculation of centrality scores is intricately tied to the eigenvector associated with the largest eigenvalue of the adjacency matrix, as large eigenvalues correspond to influential nodes. This quantification of influence helps in detecting crime leaders by spotlighting those with the most significant control over the network's dynamics.

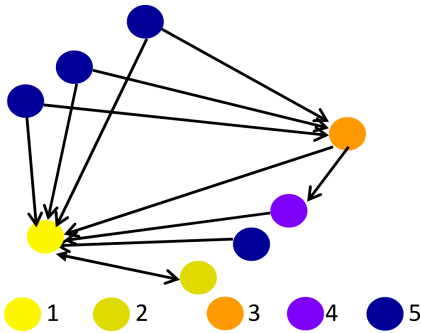

**Fig. 9:** Illustration of the procedure of PageRank. Each node's legend number indicates the centrality rank of the node.

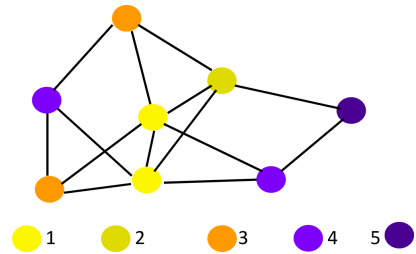

**Fig. 10:** Illustration of the procedure of Eigenvector. Each node's legend number indicates the centrality rank of the node.

*i) The Rationale Behind the Usage of the Technique*

Influential criminals may have affiliations with other influential individuals. By considering the quality and importance of these connections, eigenvector analysis can identify individuals who have significant influence. Eigenvector centrality encompasses the concept of influence diffusion, wherein the influence of a criminal is propagated through the network if they are connected to highly influential individuals. The method can uncover the hierarchical structure of influence.

*ii) The Conditions for the Technique's Optimal Performance*

To enhance the method: (1) consider directional relationships in the network. Incorporate directionality using algorithms like HITS or personalized PageRank to improve accuracy, (2) use efficient algorithms (e.g., power iteration or PageRank) to expedite the computation of eigenvector centrality for each node, (3) normalize the network size to ensure fair comparisons when comparing influences, (4) perform sensitivity analysis to evaluate the reliability of the method.

*iii) Research Papers that have Employed the Technique*

Ferrara et al. [61] introduced LogAnalysis, designed to enable the identification of criminal communities within networks created from phone call records. This system assists in comprehending the hierarchical structures within criminal organizations, uncovering influential members who facilitate connections among sub-groups. Shang and Yuan [3] conducted an assessment on the efficacy of three distinct techniques to classify an unfamiliar network into three categories: terrorist, cocaine-related, or noncriminal. The computation of eigenvector centrality for the entire network followed a similar approach as that of closeness and betweenness centralities. Calderoni et al. [38] presented a forensic system that examines the roles of individuals within a criminal organization's hierarchical structure and predicts crimes by analyzing spatiotemporal patterns of criminal activities. It utilizes eigenvector centrality to determine the significance of a node by considering the importance of its neighbors.

**Table 16:** Evaluating papers that employed Eigenvector-based analysis.

| Paper/Year | Dataset | Scalability | Interpretability | Accuracy | Efficiency |
|---|---|---|---|---|---|
| [61] 2014 | Phone log dataset | Fair | Good | Acceptable | Fair |
| [38] 2017 | Ndrangheta mafia dataset | Unsatisfactory | Acceptable | Acceptable | Unsatisfactory |

## 4.2 Role Centrality-Based Analysis

### 4.2.1 Behavior-Based Analysis

Behavior-Based Analysis represents a nuanced method specifically tailored for identifying influential criminals within social criminal networks. This approach diverges from traditional reliance on hierarchical positions or organizational roles, opting instead to meticulously observe and analyze the actual behaviors and actions of individuals within these networks. It encompasses studying their involvement in criminal activities, interactions with other network members, decision-making processes, and their capacity to persuade or influence others. By defining and calculating influence metrics, such as node importance metrics, this method measures the impact and importance of individuals, thereby pinpointing those with significant influence.

Furthermore, Behavior-Based Analysis scrutinizes the behavioral patterns of individuals, including recurring behaviors and modus operandi, while also identifying anomalies or deviations from typical behaviors within the network. This analytical perspective enables the development of predictive models aimed at forecasting future criminal activities and identifying key influencers. By focusing on the examination of behavioral aspects and employing sophisticated influence metrics, Behavior-Based Analysis effectively identifies impactful leaders within criminal networks, thereby aiding in the strategic disruption of these illicit organizations.

*i) The Rationale Behind the Usage of the Technique*

Analyzing behavior can offer valuable insights into the future conduct of individuals involved in a criminal network. Through the examination of their previous behaviors, connections, and engagements, analysts can forecast potential future criminal acts and identify emerging patterns that influential criminals might employ. Criminal networks frequently exhibit hierarchical structures. Behavior-based analysis aids in deciphering these hierarchical dynamics by observing individuals' interactions and the roles they undertake in the network. Proactive methodology empowers investigators to take preventive actions.

*ii) The Conditions for the Technique's Optimal Performance*

To improve the method: (1) access to advanced data analytics tools is crucial for handling complex data in behavior analysis (these tools should process diverse data types and detect patterns), (2) combine expertise in criminology, data analysis, and social network analysis enables pattern identification and assessment of individual influence, (3) consider geography criminal activities to provide insights into behavior patterns and network dynamics, (4) act promptly in behavior analysis to allow for better identification and response to emerging patterns.

*iii) Research Papers that have Employed the Technique*

Rodrigueza and Estuar [62] conducted a study on human behavior in disasters. They created models of perceived behavior using networks such as Agent x Agent, Agent x Knowledge, Agent x Task, and Agent x Belief. These models were analyzed across the three phases of a disaster. Additionally, they employed SNA to identify influential agents within a simulated disaster behavior network. Hutchins and Benham-Hutchins [63] conducted a study where they investigated how intelligence analysts, along with network analysis software and methodologies, utilized a combination of measures to analyze the behavior of criminal organizations. The researchers presented data from three networks to demonstrate the findings obtained through organizational risk analysis.

Easton and Karaivanov [64] explore the network structures that naturally emerge because of the interplay between a deterrence policy and the responses of networked agents as they adapt the crime network itself. The objective of the study was to gain insight into criminal behavior. Within any network, the "key player" policy identifies the individual agent whose elimination would lead to the largest decrease in total crime.

Wang et al. [65] investigated user search behavior and internet information foraging by analyzing user search sessions. The study utilized a set of search logs from a large search engine. User sessions were identified using hierarchical agglomerative clustering. Based on information foraging theory, the researchers proposed a model that predicts the probability distribution of the number of queries and clicks in a search session. The model assumes that users make sequential decisions, continuing the search if the expected value of continuing exceeds a threshold. A machine learning-driven tool for detecting and evaluating cyber threats was developed by Wang et al. [66]. The tool employs a two-stage analysis approach, incorporating both unsupervised and supervised learning techniques, and operates on a dataset of 822,226 log entries obtained from an AWS cloud-based web server. By leveraging unsupervised learning, the tool can uncover patterns, anomalies, and potential threats.

**Table 17:** Evaluating research papers that have employed Behavior-Based Analysis

| Paper/Year | Dataset | Scalability | Interpretability | Accuracy | Efficiency |
|---|---|---|---|---|---|
| [62] 2018 | Author Collected | Acceptable | Fair | Good | Good |
| [63] 2010 | Street gang data | Fair | Unsatisfactory | Good | Good |
| [65] 2007 | U.S. searches dataset | Unsatisfactory | Unsatisfactory | Acceptable | Acceptable |
| [66] 2022 | COMPAS dataset | Fair | Unsatisfactory | Fair | Acceptable |

# 5. Experimental Evaluations

Within this section, we conduct experimentation to evaluate and rank the different techniques outlined in this paper. To represent each group of algorithms that share the same underlying technique, we selected a single algorithm as a representative. Subsequently, we assessed and ranked these chosen algorithms. We ran the algorithms on a Windows 10 Pro machine, which was equipped with an Intel(R) Core(TM) i7-6820HQ processor operating at 2.70 GHz and had 16 GB of RAM.

## 5.1 Methodology for Selecting a Representative Algorithm for Each Technique and Ranking the Various Techniques

The following approach was employed for conducting the experimental evaluations:

> ➢ **Evaluating each sub-technique:** Upon conducting a thorough review of papers documenting algorithms that make use of a specific sub-technique, we successfully pinpointed the paper with the highest influence. The algorithm described in this paper was selected as the representative of the sub-technique. In order to ascertain the most noteworthy paper among all those reporting algorithms employing the same sub-technique, we evaluated several factors, including its level of innovation and publication date. Table 18 shows the list of selected papers that serve as representative of their techniques.
> ➢ **Ranking the sub-techniques that belong to the same overall technique:** We computed the average scores of the selected algorithms that employed a common sub-technique. Subsequently, we ranked the sub-techniques belonging to the same technique based on their scores.
> ➢ **Ranking the various techniques that belong to the same sub-category:** The average scores of the selected algorithms that employed a shared technique were calculated. Subsequently, we ranked the techniques falling under the same sub-category based on their scores.
> ➢ **Ranking the various sub-categories that belong to the same category:** We computed the average scores of the selected algorithms that utilized a common sub-category. We ranked the sub-categories falling under the same category based on their scores.

**Table 18:** The selected papers that serve as representative for their respective techniques.

| Technique | Paper | Technique | Paper |
|---|---|---|---|
| Katz Centrality | [20] | Spatial Random Graph Distribution | [46] |
| Multiple Link Types | [17] | Spatial Diffusion Clustering | [49] |
| Degree-Based | [67] | Density-Based Clustering | [50] |
| Closeness Centrality | [26] | Spatiotemporal Diffusion | [53] |
| Betweeness Centrality | [32] | Spatiotemporal Random Graph | [56] |
| Node Similarity-Based | [36] | PageRank-Based Analysis | [59] |
| Local Clustering Coefficient | [16] | Eigenvector-Based Analysis | [61] |
| Hierarchical Clustering | [43] | Behavior-Based Analysis | [63] |

We searched for publicly accessible codes corresponding to the algorithms we chose to represent their respective techniques. We were able to acquire codes for only the following two papers: (Cavallaro et al., [20]; Berlusconi et al., [36]). The codes for these papers are provided below:

> ➢ [20]: https://github.com/lcucav/ criminal-nets/tree/master/disruption
> ➢ [36]: https://figshare.com/articles/dataset/Oversize_network/3156067

For the remaining papers, we created our own implementations using TensorFlow, as described by Sinaga and Yang [78]. We trained these implementations using the Adam optimizer, as suggested in [78]. TensorFlow's APIs offer users the ability to develop their own algorithms (Morselli and Giguere, [79]). Python 3.6 served as our development language, and we utilized TensorFlow 2.10.0 as the models' backend.

## 5.2 Datasets

The evaluations were conducted using the following datasets:
- *Chicago Crime Dataset*: The Chicago Police Department dataset, publicly accessible and dating from 2001, details reported city crimes. Updated regularly from the Department's CLEAR system, it includes information like crime type, location, timing, and arrest records. Organized by police district, it allows for citywide crime pattern analysis and includes fields like dates, addresses, coordinates, FBI codes, and location types. The dataset is downloadable as the Chicago Crime Dataset [72].
- *San Francisco crime dataset*: The dataset from San Francisco, featuring 39 crime categories, shows larceny/theft as the most common. The dataset was sourced from the San Francisco crime dataset [73].

## 5.3 Evaluation Setup

Common Parameters:
- Maximum Number of Iterations (Katz centrality [20] and Eigenvector-Based Analysis [61]): Set to 100.
- Minimum Cluster Size (Hierarchical-Based Clustering [43], Spatial Random Graph Distribution-Based Clustering [46], and Spatiotemporal Diffusion-Based Analysis [49]): Set to 30.
- Similarity/Dissimilarity Thresholds (Node Similarity-Based Model [36], and Hierarchical-Based Clustering [43]): Set to 0.5.
- Random Seed (Spatial Random Graph [46] and Spatiotemporal Random Graph [56]): Set to 42.
- Termination/Convergence Threshold (Katz Centrality [20], PageRank-Based Analysis [59], and Eigenvector-Based Analysis [61]): Set to 1e-4 or 0.0001.

Unique Parameters:
- Katz centrality [20]: We assigned a value of α = 0.1 to the attenuation factor (α), which regulates the impact of remote nodes on the centrality score.
- Node Similarity-Based Model [36]: We took the following into consideration: (1) we employed Cosine Similarity as the metric to measure similarity, (2) TF-IDF vectors were used to represent the features, (3) for each node, we identified the top 4 nearest neighbors, and (4) we applied a similarity threshold of 0.5 to remove connections that were considered weak.
- Local Clustering Coefficient-Based Model [16]: For neighborhood size, we considered a larger neighborhood (degree-3 neighbors). We considered 0.001 for Learning Rate, 0.01 for Regularization Strength, 2 for number of layers, 64 for number of hidden units, 32 for batch size, 0.5 for dropout rate, and 100 for number of epochs.
- Closeness [26] and Betweeness Centralities [32]: We utilized the conventional closeness centrality metric, which accounts for the shortest distance between nodes. We disregarded nodes that cannot be reached from other nodes. Then, we standardized the closeness centrality scores within the range of 0 to 1.
- Hierarchical-Based Clustering [43]: We employed the average linkage criterion and Euclidean distance as the distance metric for our hierarchical-based clustering. To achieve the desired number of clusters, we opted to truncate the dendrogram at a fusion coefficient of 0.5. Additionally, we established a distance threshold of 0.5, whereby the merging process halts if the dissimilarity between two clusters exceeds this value. We imposed a minimum cluster size of 50 data points, terminating the merging if a cluster falls below this threshold. We set a termination threshold of 5%, causing the algorithm to cease merging if the dissimilarity between two merge steps decreases by less than this percentage.
- Spatial Random Graph Distribution-Based Clustering [46]: We have defined the following parameters for our clustering algorithm: (1) Radius (R): We have set R to 100 meters, which determines the spatial proximity of points in the dataset, (2) k-nearest neighbors (k): We have set k to 5, which determines the number of nearest neighbors considered when constructing the spatial graph, (3) Distribution Threshold (DT): We have set DT to 0.6, which determines the threshold value used to decide if an edge should exist between two data points based on their distribution similarity, (4) Minimum Cluster Size (min_cluster_size): We have set it to 10, specifying the minimum number of points required for a cluster to be considered valid, (5) Spatial Density Threshold (SDT): We have set SDT to 0.4, which determines the threshold used to determine if a cluster is spatially dense enough, and (6) Seed: We set the seed to 42, which ensures the reproducibility of the clustering results by using a specific random seed value.
- Spatial Diffusion-Based Clustering [49]: We establish these following values: (1) DT = 0.1 as the Distance Threshold (DT), dictating the maximum allowable distance between two data points to qualify them as neighbors, (2) TT = 5 minutes as the Time Threshold (TT), the maximum time difference between two data points to classify them as neighbors, 5 as the Minimum Cluster Size, and DR = 0.5 as the Medium Diffusion Rate.
- Density-Based Clustering [50]: We establish the following values: (1) ε = 0.5 as the Epsilon (ε), indicating the radius within which the algorithm scans for neighboring points, and (2) MinPts = 5 as the Minimum Points (MinPts), defining the minimum number of points needed within the ε radius to constitute a dense region or cluster.
- Spatiotemporal Diffusion-Based Analysis [53]: We have defined the following parameters for our analysis: (1) Time Step (Δt): We have set Δt to be 1 hour. This value determines the duration between consecutive time points in the analysis, (2) Spatial Step (Δx, Δy, Δz): The values of Δx, Δy, and Δz have been set to 1 kilometer each. These values determine the distance between adjacent spatial grid points in each dimension, (3) Diffusion Coefficient (D): The value of D is 0.1 square kilometer per hour. This coefficient measures the speed at which the diffusing substance spreads and determines the rate of diffusion, (4) The initial concentration (C₀) of the diffusing substance at the beginning of the analysis is set to C₀ = 1.
- PageRank-Based Analysis [59]: We used the following settings: (1) damping factor (d): We chose a value of 0.9 for the damping factor. This factor determines the likelihood that a random surfer will continue clicking on links instead of jumping to a random node, and (2) convergence threshold for PageRank scores: We specified a small change in PageRank scores as the convergence threshold (to determine when the PageRank has reached a stable converged).

## 5.3 Model for Simulating the Spreading Ability and Metrics for Evaluating the Performance of the Algorithms

To assess the accuracy of the ranked list of influential nodes generated by an algorithm, we conducted a comparison between the list and an actual propagation process involving the nodes. This evaluation was carried out using a widely recognized procedure outlined in Chen et al. [74]:

i. Recording the list of nodes ranked by each algorithm.

ii. Utilizing the SIR model (Chen et al. [74]) to simulate the spreading ability of nodes. Within this model, every node is assigned to one of three states: susceptible, infected, or recovered. In each state, only one node is considered infected. Infected nodes can infect their susceptible neighbors with a specific probability of spreading. In the experiments, we set the spreading probability $\beta$ = 0.1-0.2, and the recovery probability $\mu$ = 1. Initially, we set the top-$k$ ranked nodes to be infected, where $k$ = 1% * $n$ ($n$ is the number of nodes). Then, the number of infected nodes increases based on the SIR model.

iii. Using the nodes ranked by one of the algorithms and the corresponding one ranked by the SIR model, we recorded the pair scores in a list. This survey utilizes the following three metrics to assess the performance of the algorithms:

- *Kendall's tau correlation coefficient (Kendall [75])*: It gauges the resemblance of data rankings between two quantities. The value of $\tau$ is in the range {+1, -1}. It is defined as shown in Equation 1.

$$\tau = \frac{N_1 - N_2}{0.5 N (N-1)} \quad (1)$$

where $N_1$ and $N_2$ are the number of concordant and discordant pairs, respectively.

- *Monotonicity index (Zareie [76])*: It is a metric used for quantifying the resolution of different indices. It is defined as shown in Equation 2:

$$M(L) = \left[ 1 - \frac{\sum_{l \in L} |V|_l * (|V|_l - 1)}{|V| * (|V| - 1)} \right]^2 \quad (2)$$

where $|V|_l$ is the number of nodes that have the same index rank $l$ in the ranked list $L$; and $V$ is the number of nodes.

- *Percentage average absolute error (Saxena [77])*: It is a numerical amount of the discrepancy between an exact value and the corresponding estimated one. The absolute error (AE($v$)) for a node $v$ is defined as in Equation 3:

$$AE(v) = |EST(v) - ACT(v)| \quad (3)$$

The percentage average absolute error PAAE($v$) for the node v is defined as shown in Equation 4:

$$PAAE(v) = \frac{\text{Average absolute error}}{\text{network size}} 100\%$$

## 5.4 The Experimental Results

Tables 18-20 and Figs. 11-13 present the experimental results.

**Table 18:** Kendall's Coefficient scores of the algorithms. The table also shows the following rankings: (1) the various sub-techniques that belong to the same technique, (2) the various techniques that belong to the same sub-category, (3) the various sub-categories that belong to the same category, and (4) the categories.

| Cat. | Sub-Cat. | Technique | Sub-Technique | Selected Papers | Data sets | Score | Sub-Tech. Rank | Tech Rank | Sub-Cat. Rank | Cat. Rank |
|---|---|---|---|---|---|---|---|---|---|---|
| Topology-Based | Global Analysis | Network-Based Analysis | Katz Centrality | Cavallaro [20] | Chi / SFO | 0.851 / 0.858 | 1 | 2 | 1 | 1 |
| | | | Multiple Link Types | Bright [17] | Chi / SFO | 0.746 / 0.774 | 2 | | | |
| | | | Degree-Based | Bright [67] | Chi / SFO | 0.674 / 0.698 | 3 | | | |
| | | Shortest Path Analysis | Closeness Centrality | Calderon [26] | Chi / SFO | 0.819 / 0.826 | 2 | 1 | | |
| | | | Betweeness Centrality | Taha [32] | Chi / SFO | 0.829 / 0.832 | 1 | | | |
| | Local Analysis | Node Similarity-Based | N/A | Berlusconi [36] | Chi / SFO | 0.804 / 0.815 | N/A | 3 | 2 | |
| | | Local Clustering Coefficient | N/A | Agreste [16] | Chi / SFO | 0.784 / 0.792 | N/A | 4 | | |
| Clustering-Based | Spatial Analysis | Global Analysis | Hierarchical Analysis | Xu [43] | Chi / SFO | 0.772 / 0.783 | 1 | 1 | 1 | 3 |
| | | | Spatial Random G. Distribution | Agarwal [46] | Chi / SFO | 0.668 / 0.674 | 2 | | | |
| | | Local Analysis | Spatial Diffusion Clustering | Taha [49] | Chi / SFO | 0.665 / 0.708 | 2 | 2 | | |
| | | | Density-Based Clustering | Everton [50] | Chi / SFO | 0.687 / 0.710 | 1 | | | |
| | Spatio-Temporal | Spatiotemporal Diffusion | N/A | Zhao [53] | Chi / SFO | 0.649 / 0.693 | N/A | 1 | 2 | |
| | | Spatiotemporal Random G. | N/A | Berlusconi [56] | Chi / SFO | 0.637 / 0.644 | N/A | 2 | | |
| Agent-Based | Eigenvector Centrality | PageRank-Based | N/A | Isah [59] | Chi / SFO | 0.882 / 0.911 | N/A | 1 | 1 | 2 |
| | | Eigenvector | N/A | Ferrara [61] | Chi / SFO | 0.879 / 0.899 | N/A | 2 | | |
| | Role Based | Behavior-Based | N/A | Hutchins [63] | Chi / SFO | 0.611 / 0.623 | N/A | 1 | 2 | |

**Table 19:** Monotonicity scores of the algorithms. The table also shows the following rankings: (1) the various sub-techniques that belong to the same technique, (2) the various techniques that belong to the same sub-category, (3) the various sub-categories that belong to the same category, and (4) the various categories.

| Cat. | Sub-Cat. | Technique | Sub-Technique | Selected Papers | Data sets | Score | Sub-Tech. Rank | Tech Rank | Sub-Cat. Rank | Cat. Rank |
|---|---|---|---|---|---|---|---|---|---|---|
| Topology-Based Analysis | Global Analysis | Network-Based Analysis | Katz Centrality | Cavallaro [20] | Chi / SFO | 0.992 / 0.994 | 1 | 1 | 1 | 1 |
| | | | Multiple Link Types | Bright [17] | Chi / SFO | 0.989 / 0.992 | 2 | | | |
| | | | Degree-Based | Bright [67] | Chi / SFO | 0.957 / 0.962 | 3 | | | |
| | | Shortest Path Analysis | Closeness Centrality | Calderon [26] | Chi / SFO | 0.991 / 0.990 | 1 | 2 | | |
| | | | Betweeness Centrality | Taha [32] | Chi / SFO | 0.984 / 0.988 | 2 | | | |
| | Local Analysis | Node Similarity-Based | N/A | Berlusconi [36] | Chi / SFO | 0.981 / 0.985 | N/A | 1 | 2 | |
| | | Local Clustering Coefficient | N/A | Agreste [16] | Chi / SFO | 0.977 / 0.982 | N/A | 2 | | |
| Clustering-Based | Spatial Analysis | Global Analysis | Hierarchical Analysis | Xu [43] | Chi / SFO | 0.965 / 0.973 | 1 | 1 | 1 | 3 |
| | | | Spatial Random G. Distribution | Agarwal [46] | Chi / SFO | 0.961 / 0.969 | 2 | | | |
| | | Local Analysis | Spatial Diffusion Clustering | Taha [49] | Chi / SFO | 0.962 / 0.970 | 1 | 2 | | |
| | | | Density-Based Clustering | Everton [50] | Chi / SFO | 0.958 / 0.967 | 2 | | | |
| | Spatio-Temporal | Spatiotemporal Diffusion | N/A | Zhao [53] | Chi / SFO | 0.951 / 0.968 | N/A | 2 | 2 | |
| | | Spatiotemporal Random G. | N/A | Berlusconi [56] | Chi / SFO | 0.955 / 0.963 | N/A | 1 | | |
| Agent-based | Eigenvector Centrality | PageRank-Based | N/A | Isah [59] | Chi / SFO | 0.992 / 0.995 | N/A | 1 | 1 | 2 |
| | | Eigenvector | N/A | Ferrara [61] | Chi / SFO | 0.983 / 0.987 | N/A | 2 | | |
| | Role Based | Behavior-Based | N/A | Hutchins [63] | Chi / SFO | 0.972 / 0.973 | N/A | N/A | 2 | |

**Table 20:** % average absolute error scores of the algorithms. The table also shows the following rankings: (1) the various sub-techniques that belong to the same technique, (2) the various techniques that belong to the same sub-category, (3) the various sub-categories that belong to the same category, and (4) the categories.

| Cat. | Sub-Cat. | Technique | Sub-Technique | Selected Papers | Data sets | Score | Sub-Tech. Rank | Tech Rank | Sub-Cat. Rank | Cat. Rank |
|---|---|---|---|---|---|---|---|---|---|---|
| Topology-Based Analysis | Global Analysis | Network-Based Analysis | Katz Centrality | Cavallaro [20] | Chi / SFO | 0.783 / 0.537 | 1 | 1 | 1 | 2 |
| | | | Multiple Link Types | Bright [17] | Chi / SFO | 0.847 / 0.658 | 2 | | | |
| | | | Degree-Based | Bright [67] | Chi / SFO | 0.616 / 0.409 | 3 | | | |
| | | Shortest Path Analysis | Closeness Centrality | Calderon [26] | Chi / SFO | 1.573 / 0.873 | 2 | 2 | | |
| | | | Betweeness Centrality | Taha [32] | Chi / SFO | 1.097 / 0.724 | 1 | | | |
| | Local Analysis | Node Similarity-Based | N/A | Berlusconi [36] | Chi / SFO | 2.107 / 0.566 | N/A | 1 | 2 | |
| | | Local Clustering Coefficient | N/A | Agreste [16] | Chi / SFO | 2.647 / 0.941 | N/A | 2 | | |
| Clustering-Based | Spatial Analysis | Global Analysis | Hierarchical Analysis | Xu [43] | Chi / SFO | 4.860 / 1.467 | 1 | 1 | 2 | 3 |
| | | | Spatial Random G. Distribution | Agarwal [46] | Chi / SFO | 4.491 / 1.142 | 2 | | | |
| | | Local Analysis | Spatial Diffusion Clustering | Taha [49] | Chi / SFO | 5.853 / 2.073 | 1 | 2 | | |
| | | | Density-Based Clustering | Everton [50] | Chi / SFO | 5.907 / 2.651 | 2 | | | |
| | Spatio-Temporal | Spatiotemporal Diffusion | N/A | Zhao [53] | Chi / SFO | 2.964 / 3.132 | N/A | 1 | 1 | |
| | | Spatiotemporal Random G. | N/A | Berlusconi [56] | Chi / SFO | 3.482 / 3.292 | N/A | 2 | | |
| Agent-based | Eigenvector Centrality | PageRank-Based | N/A | Isah [59] | Chi / SFO | 0.408 / 0.325 | N/A | 1 | 1 | 1 |
| | | Eigenvector | N/A | Ferrara [61] | Chi / SFO | 0.432 / 0.338 | N/A | 2 | | |
| | Role Based | Behavior-Based | N/A | Hutchins [63] | Chi / SFO | 0.695 / 0.454 | N/A | N/A | 2 | |

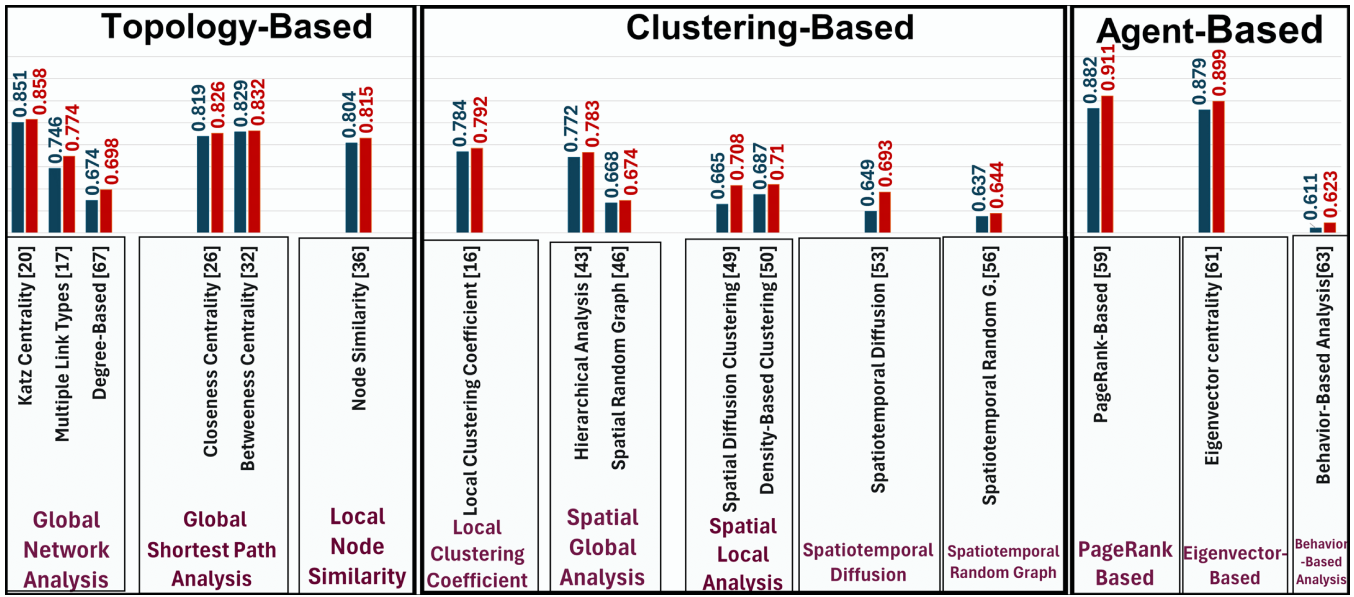

**Fig. 11.** Kendall's tau correlation coefficient $\tau$ scores of the algorithms. The algorithms are grouped based on the common underlying techniques they employ.

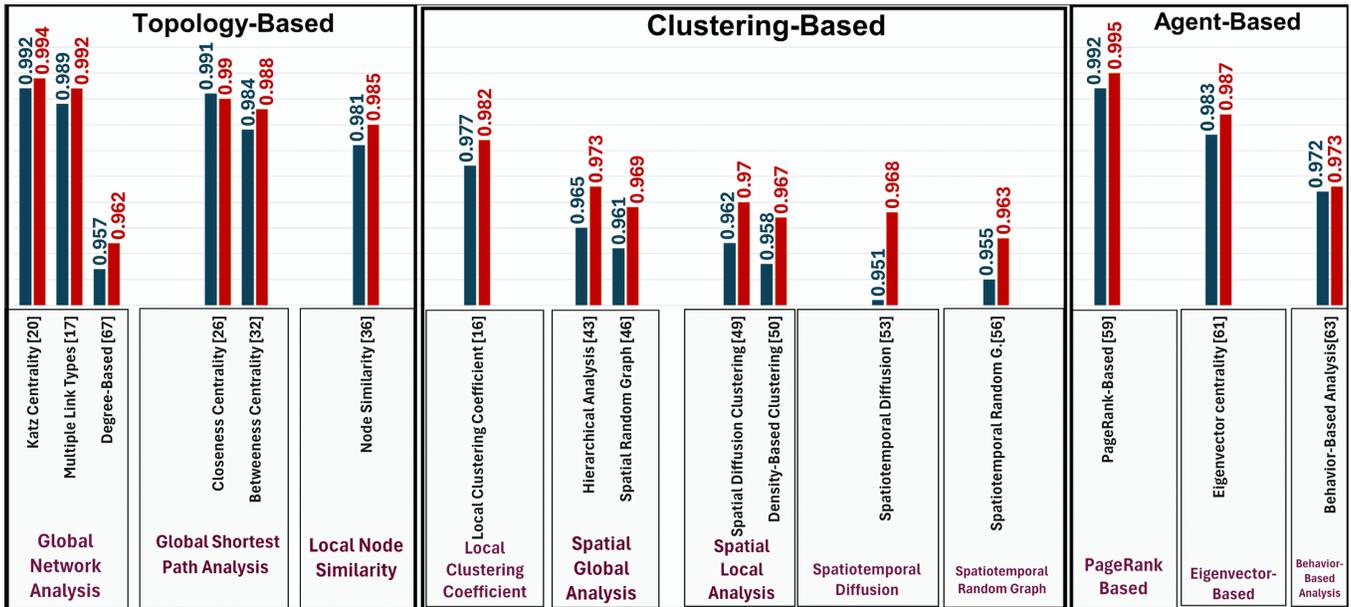

**Fig. 12.** Monotonicity scores of the selected algorithms. The algorithms are grouped based on the common underlying techniques they employ.

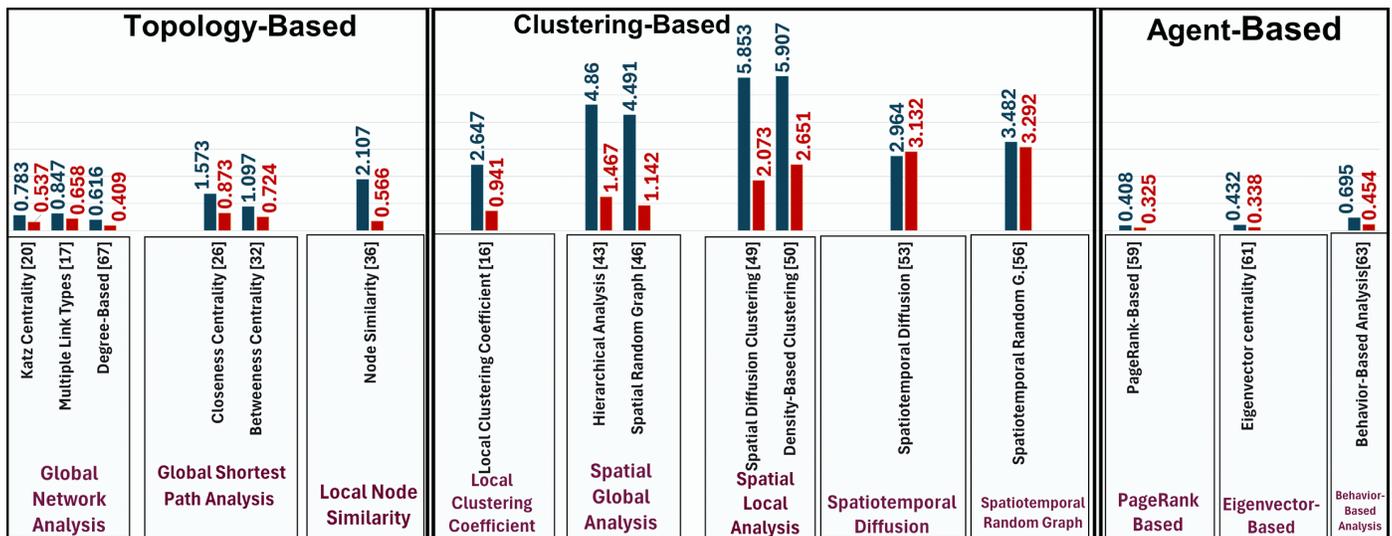

**Fig. 13.** % average absolute error scores of the selected algorithms. The algorithms are grouped based on the common underlying techniques they employ.

## 5.5 Discussion of the Experimental Results

### 5.5.1 Shortest Path-Based Analysis (Closeness-Based Centrality and Betweeness-Based Centrality)

The algorithms applied to the SIR model performed well overall. Algorithms based on betweenness centrality, however, were less effective in networks where information doesn't follow the shortest paths and failed to predict the influence of low-connectivity nodes. The inclusion of weak ties decreased accuracy, especially during initial stages, but performance improved in later stages and in well-connected networks. Closeness centrality-based algorithms efficiently computed node rankings, with a sigmoid pattern observed between closeness centrality and reverse ranking. Minor edge additions significantly boosted node centrality and ranking, identifying potential key players in criminal networks due to their proximity to other nodes. Algorithms using betweenness centrality excelled in blocking information flow and identifying influential nodes, especially when considering the ego-betweenness of top-ranked nodes. These results were efficient and comparable to general betweenness outcomes. This approach emphasized the role of individual connections in linking subgroups and spreading information, showing that individuals with high betweenness centrality are often involved in various criminal activities across different subgroups.

### 5.5.2 Spatial Global Analysis-Based (Spatial Random Graph and Distribution, and Hierarchical Clustering)

The tested algorithms effectively propagated through networks, outperforming the single-point contact SIR model in generating infected nodes and identifying top-10 nodes with high accuracy. Random-based filtering in these algorithms showed superior performance compared to other approaches using a similar strategy, particularly effective in selecting high-degree nodes as seed nodes. However, instability issues arose due to the randomness in selection. The algorithms excelled in large-scale networks, and experiments varying the parameters of the spatial random graph distribution algorithm indicated robust performance in identifying crime leaders. This approach, emphasizing geographic proximity, was more effective in pinpointing crime leaders by leveraging spatial patterns in criminal activities and social connections. The influential nodes identified by these algorithms had a greater impact on spreading information compared to those ranked highly by betweenness, eigenvector, or PageRank metrics. In terms of efficiency, these algorithms surpassed many popular Influence Maximization algorithms.

### 5.5.3 Spatiotemporal Clustering (Spatiotemporal Diffusion and Spatiotemporal Random Graph)

The node ranking correlation between the SIR model and the algorithms showed moderate agreement, with modest accuracy across different $\beta$ values. The algorithms performed slightly better with $\beta$ between 0.13 and 0.15, but less effectively for the top 3% of nodes. However, they successfully identified influential low-degree nodes near network cores. As the number of core nodes per community was adjusted, the algorithms' performance approached that of the SIR model, suggesting potential for optimizing diffusion performance. The Spatiotemporal Diffusion algorithms revealed temporal trends and identified influential individuals during specific periods, aiding law enforcement in resource allocation and targeting. By incorporating spatial data, these algorithms pinpointed crime clusters and key local nodes, suggesting strategies for disrupting localized criminal networks. Similarly, Spatiotemporal Random Graph algorithms detected temporal shifts in node centrality, highlighting nodes with fluctuating influence and bridge nodes connecting different network areas. Monitoring these nodes could disrupt cross-regional criminal activities and limit crime spread. These findings offer strategic insights for law enforcement in tackling criminal networks.

### 5.5.4 Eigen centrality Analysis (Eigenvector and PageRank)

The algorithms effectively handled varying propagation probabilities in networks, excelling in identifying structurally important nodes. They captured local dynamics through bridge-like structures and evaluated global roles based on key bridge connections. Infection extent depended on a node's neighbor count and their propagation capabilities, efficiently pinpointing central nodes in clusters and bridge nodes. In weighted social networks, these algorithms surpassed other centrality measures, highlighting an inverse relationship between sub-graph density and node centrality. They remained stable against random network perturbations and excelled in distinguishing nodes with different spreading abilities, balancing accuracy with computational efficiency. Eigenvector-Based algorithms effectively identified influential individuals in criminal networks, distinguishing crime leaders with high centrality scores. These leaders, often involved in multiple criminal activities, formed influence clusters within the network, with the algorithms accurately capturing the hierarchical structure. PageRank-Based algorithms also successfully identified key crime leaders by assessing interconnectedness and influence. They excelled in ranking individuals by applying PageRank principles to analyze criminal network dynamics.

### 5.5.5 Network-Based Model Analysis (Katz Centrality, Multiple Link Types, and Degree Centrality)

The analysis revealed that individuals with high Katz centrality scores, indicating many direct and indirect connections, are potential criminal leaders. This centrality measure identified influential individuals who may not have numerous direct connections but are strongly connected to other central figures, underscoring the role of indirect connections in understanding influence within the criminal network. High Katz centrality scores also helped uncover cohesive clusters or subgroups, representing distinct criminal organizations or factions led by central figures. The Multiple Link Types Model further differentiated the roles of criminal leaders based on various relationship types, like co-offending, communication, and financial transactions. For instance, high centrality from co-offending relationships indicated an individual's ability to coordinate criminal activities, while centrality from communication patterns pointed to their role as information hubs. Financial transaction-based centrality highlighted control over criminal finances. By analyzing individuals with high overall centrality scores in this model, key subgroups or clusters were identified, representing different criminal factions led by central figures. This approach provided valuable insights into the structure and dynamics of the criminal social network. Experimentally, the Degree-Based centrality algorithm showed some success in identifying crime leaders by focusing on individuals with numerous connections, yet it only achieved moderate performance, suggesting a limited ability to pinpoint true crime leaders. The behavior-based centrality algorithm, on the other hand, effectively identified crime leaders by combining behavioral attributes with network structure, distinguishing leaders from other network members more accurately.

### 5.5.6 Spatial Local Analysis (Density-Based Clustering, Fuzzy C-Means Clustering, and Spatial diffusion Clustering)

By applying the Density-Based Clustering (DBSCAN) algorithm with optimized parameters to the criminal social network dataset, clusters representing various criminal groups were identified. The algorithm effectively pinpointed potential crime leaders in each cluster based on their network positions and connections. The DBSCAN algorithm showed proficiency in identifying crime leaders, as evidenced by the strong cohesion within clusters and clear separation between them, indicated by the silhouette coefficient. Similarly, the Fuzzy C-Means (FCM) algorithm demonstrated effective performance in identifying crime leaders within the same network. The algorithm's membership values quantified individuals' associations with different criminal groups, facilitating the identification of influential members. Additionally, centrality measures indicated that these identified crime leaders held prominent positions within the network, exerting significant influence over other members.

### 5.5.7 Role centrality Analysis (Behavior-Based Analysis)

This algorithm, focusing on local structural information, failed to consider the global network structure, leading to the identification of high-degree but low-influence nodes. Its accuracy declined with increased spreading degrees, performing poorly at 4-hop spreading. As the value of k increased, the propagation capability of the top-k nodes diminished. The algorithm's performance worsened with larger datasets, particularly in networks with nodes of smaller degrees. In networks with unclear community structures, the algorithm underperformed, though it had acceptable propagation range and transmission rates in other networks. It efficiently summarized node connectivity without needing to analyze the entire network topology.

## 6. Potential Future Perspectives on Techniques for Identifying Crime Leaders

### 6.1 Katz Centrality

Katz centrality measures the influence of a node based on the number and importance of its neighboring nodes. In the future, Katz centrality could be enhanced by incorporating additional factors such as temporal dynamics, sentiment analysis, or multi-layer networks. By considering the evolving nature of criminal activities, sentiment analysis can help identify key individuals who exhibit patterns of involvement or influential behavior.

### 6.2 Multiple Link Types

Criminal social networks often have different types of relationships between individuals. Extending centrality measures to accommodate multiple link types could provide more nuanced insights. For example, a criminal network may have connections based on financial transactions, personal relationships, or shared locations. Incorporating such diverse link types into centrality calculations can reveal different dimensions of influence and identify individuals with varying roles in criminal activities.

### 6.3 Closeness Centrality

Closeness centrality quantifies the accessibility of a node within a network, based on the shortest paths to other nodes. Future perspectives for closeness centrality in criminal social networks could involve incorporating geographic factors, such as proximity to crime scenes or hotspots, as well as time-dependent factors, such as the frequency of interactions. By considering spatiotemporal aspects, closeness centrality can identify influential criminals who are geographically well-positioned and actively engaged in criminal activities.

### 6.4 Betweenness Centrality

Betweenness centrality measures the extent to which a node lies on the shortest paths between other nodes. In the future, betweenness centrality could be enhanced by considering the context of criminal activities and the flow of information or resources within the network. By incorporating additional information, such as the nature of criminal transactions or the exchange of illegal goods, betweenness centrality can identify individuals who act as intermediaries, controlling the flow of resources, or those who bridge groups within the network.

### 6.5 Node Similarity-Based Analysis

Future advancements in node similarity-based analysis could integrate graph neural networks (GNNs) to better capture complex patterns in criminal networks. GNNs can learn rich node embeddings, considering nodes and their neighborhoods, thus enhancing the identification of influential criminals by their similarity to known influential individuals.

### 6.6 Local Clustering Coefficient

Future use of the local clustering coefficient in criminal networks could include dynamic aspects of criminal activities, focusing on temporal patterns where clusters form or dissolve over time. Analyzing evolving local clustering coefficients can help identify influential criminals central to these temporal cluster changes.

### 6.7 Hierarchical Analysis

Hierarchical analysis aims to identify hierarchical structures within a network, such as nested clusters or levels of influence. In the future, advancements in hierarchical analysis for identifying influential criminals in a criminal social network could involve the integration of multi-resolution techniques. These techniques would allow for the identification of influential individuals at different scales, capturing both macro-level structures and micro-level dynamics. By uncovering hierarchical patterns of influence, law enforcement agencies can better understand the organization and power dynamics within criminal networks.

### 6.8 Spatial Random Graph Distribution

In the future, spatial random graph distribution for identifying influential criminals could integrate geographic information systems (GIS) and spatial analytics. This approach would analyze the distribution of criminal activities and individual connectivity within geographical areas, identifying criminals with significant local presence or those linking different regions. Combining spatial random graph distribution with GIS can offer insights into the dynamics of criminal networks.

### 6.9 Spatial Diffusion Clustering

Spatial diffusion clustering focuses on the spread of information or activities within a network over space and time. Future perspectives for this approach in identifying influential criminals could involve the incorporation of machine learning algorithms capable of modeling and predicting the spatial diffusion of criminal activities. By understanding the patterns of how criminal activities spread and identifying individuals who are central to these diffusion processes, law enforcement agencies can effectively target and disrupt criminal networks.

### 6.10 Density-Based Clustering

Density-based clustering algorithms aim to identify clusters in a network based on the density of connections. In the context of identifying influential criminals, future perspectives for density-based clustering could involve the integration of multiple data sources, such as social media data, telecommunications records, or financial transactions. By combining network data with external data sources, law enforcement agencies can gain a more comprehensive understanding of criminal networks and identify influential individuals based on their connectivity patterns and the richness of information available.

### 6.11 Spatiotemporal Diffusion

Spatiotemporal diffusion analysis focuses on understanding the spread of information, behaviors, or activities over both space and time. In the context of identifying influential criminals, future perspectives for spatiotemporal diffusion analysis could involve the integration of advanced machine learning techniques and big data analytics, as follows:

1. *Predictive Modeling:* Machine learning advancements, particularly spatiotemporal forecasting models, enable prediction of future criminal activity using historical data. These models consider factors like spatial proximity, temporal patterns, social dynamics, and environmental influences, helping law enforcement identify and prioritize individuals likely to be central in criminal activities.
2. *Real-Time Monitoring:* Incorporating real-time data like social media, surveillance footage, and sensor data improves spatiotemporal diffusion analysis. Law enforcement can use this for real-time monitoring of criminal activities, identifying key players in criminal operations and the diffusion process. This enables proactive disruption of criminal networks.

### 6.12 Spatiotemporal Random Graph

Spatiotemporal random graph analysis considers both the spatial and temporal dimensions of a network, incorporating the interactions between nodes over time and across geographical locations. Future perspectives for spatiotemporal random graph in identifying influential criminals could involve the following:

1. *Network Evolution Analysis:* Criminal networks, characterized by evolving strategies, alliances, and activities, change over time. Analyzing their spatiotemporal evolution helps identify criminals with stable influence or those rapidly gaining power. Understanding these dynamics offers insights into the stability and resilience of criminal organizations.
2. *Community Detection:* Community detection algorithms in spatiotemporal random graphs reveal clusters of individuals with strong spatiotemporal connections. Identifying these densely connected criminal groups enables law enforcement to disrupt these communities and target influential criminals linking different clusters.

### 6.13 PageRank-Based Analysis

PageRank, which gauges node importance in a network through connectivity and neighbor importance, can be enhanced for identifying influential criminals by incorporating temporal dynamics, criminal activity patterns, and individual attributes. Adapting PageRank-based algorithms to consider the evolving nature of criminal networks and individual behaviors over time can more accurately capture the influence of criminals who change strategies or engagement patterns.

### 6.14 Eigenvector Centrality

Eigenvector centrality, which assesses a node's influence based on its connections and its neighbors' influence, can be improved for identifying influential criminals by integrating multi-layer or multi-modal networks. Considering various types of interactions in criminal networks, like financial transactions, communication, or shared locations, will allow eigenvector centrality to offer a more holistic assessment of criminal influence across multiple dimensions.

### 6.15 Behavior-Based Analysis

Future approaches to behavior-based analysis for identifying influential criminals could integrate machine learning to analyze complex behavioral patterns. Leveraging advanced analytics, like anomaly detection algorithms, this method can pinpoint influential criminals with unique or abnormal behaviors, uncovering hidden key individuals not easily detected through network structure alone.

### 6.16 Degree Centrality

Future applications of degree centrality in identifying influential criminals could involve using weighted networks to account for the strength or importance of connections. This approach would allow for a more nuanced understanding of influence by considering the intensity or significance of relationships, helping to identify criminals with not only numerous connections but also influential ties in the network.

## 7. Conclusions

This survey paper tackles the issue of vague and generalized categorizations in algorithmic approaches to crime leader identification and prediction. Traditional surveys often use broad classifications, leading to misalignments and imprecise evaluations. In response, our work introduces a novel, detailed methodological taxonomy, specifically for predicting crime leaders. We divide crime leader identification algorithms into three main classes: topology-based, clustering-based, and agent-based methods. Each class is further subdivided into three increasingly specific tiers, refining categorization and improving the precision and assessment of algorithms. Our key contributions are threefold as followed:

1. Our survey provides a detailed analysis of crime prediction algorithms, focusing on their sub-techniques, techniques, and categories. This taxonomy aids in accurate assessments, enhancing understanding of these algorithms' strengths and limitations, crucial for future research.
2. We conducted an empirical evaluation of techniques for identifying crime leaders, using four criteria to offer insights into their practical efficacy and applicability.
3. Our experimental evaluation compares and ranks numerous algorithms across different levels: sub-techniques, techniques, sub-categories, and categories. This comprehensive analysis gives a nuanced view of their performance and appropriateness in various scenarios.

Below, we present the main discoveries from our experimental outcomes:

- *The techniques yielded the best results (PageRank-Based and Eigenvector):* The algorithms excelled across varying propagation probabilities, effectively mapping structural dependencies in dense network regions. They proficiently identified both central nodes in clusters and connectors between network segments. An inverse relationship was noted between sub-graph density and node centrality. Resilient to random network perturbations, these algorithms surpassed others in assigning distinct rankings to nodes based on their spreading capabilities, balancing sorting accuracy with computational efficiency. Their accurate portrayal of hierarchical structures in networks highlights their potential in pinpointing key players and provides deep insights into the structural aspects of criminal networks.
- *The technique that achieved the second highest performance (Katz Centrality and Multiple Link Types):* Katz centrality effectively identified influential individuals in a criminal network, emphasizing the role of indirect connections. This measure revealed distinct clusters or subgroups, likely representing different criminal organizations or factions led by central figures, demonstrating its ability to uncover the network's hierarchical structure. The model highlighted the roles and characteristics of criminal leaders. By analyzing individuals with high centrality scores in this model, key subgroups or clusters associated with various criminal organizations were identified, further indicating the presence of central figures or leaders in these groups.
- *The technique that achieved the lowest performance (Spatiotemporal Random Graph):* The agreement between node rankings in the SIR model and the algorithms was moderate, varying with different *β* values, indicating modest accuracy in node ranking. As the number of ranked nodes increased, the algorithms' ability to improve rankings diminished. While there was a slight improvement within the β range of 0.13 to 0.15, the algorithms performed poorly among the top 3% of ranked nodes.